\documentclass[twocolumn, superscriptaddress, floatfix, longbibliography, nofootinbib]{revtex4-1}

\usepackage{amsmath}
\usepackage{graphicx}
\usepackage{amssymb}
\usepackage{mathtools} 
\usepackage[colorlinks=true, urlcolor=blue, anchorcolor=blue, citecolor=blue, linkcolor=blue]{hyperref}

\begin{document}

\author{Evangelos S. Papaefthymiou}
\author{Costas Iordanou}
\author{Fragkiskos Papadopoulos}
\email{f.papadopoulos@cut.ac.cy}
\affiliation{Department of Electrical Engineering, Computer Engineering and Informatics, Cyprus University of Technology, 3036 Limassol, Cyprus}
\date{\today}

\title{Fundamental dynamics of popularity-similarity trajectories in real networks}

\begin{abstract}
Real networks are complex dynamical systems, evolving over time with the addition and deletion of nodes and links. Currently, there exists no principled mathematical theory for their dynamics---a grand-challenge open problem. Here, we show that the popularity and similarity trajectories of nodes in hyperbolic embeddings of different real networks manifest universal self-similar properties with typical Hurst exponents $H \ll 0.5$. This means that the trajectories are predictable, displaying anti-persistent or `mean-reverting' behavior, and they can be adequately captured by a fractional Brownian motion process. The observed behavior can be qualitatively reproduced in synthetic networks that possess a latent geometric space, but not in networks that lack such space, suggesting that the observed subdiffusive dynamics are inherently linked to the hidden geometry of real networks. These results set the foundations for rigorous mathematical machinery for describing and predicting real network dynamics.
\end{abstract}

\maketitle

Modeling and prediction of network dynamics, i.e., of the connections and disconnections that take place in a given network over different time scales, is perhaps the most fundamental unresolved problem in complex networks~\cite{brain_theory_2014}, listed also in the popular 23 Mathematical Challenges of DARPA~\cite{DARPA}. Accurate prediction of such dynamics can unlock a plethora of applications across diverse domains, ranging from early detection and prevention of critical events in networked systems, to decision and policy-making in financial markets, to strategic interventions for counter-terrorism and public health~\cite{easley2010, Gutfraind2009}. 
 
We observe that the mathematical machinery and equations that describe the evolution of other complex dynamical systems and temporal data, such as gravitational and molecular systems, turbulence, and financial data, have been well-developed and known for decades~\cite{alter1959, aarseth2003, schlick2010molecular, friedrich2011, Mandelbrot1968}. Yet, the underlying processes governing network dynamics and their mathematics remain elusive. One of the main reasons for this discrepancy is the fact that networks are discrete topological structures, not inheriting the standard form of temporal data met in other classical dynamical systems.  

Advancements in network geometry during the past years revealed that real networks can be meaningfully mapped into continuous hyperbolic spaces~\cite{Krioukov2010, Boguna2010, Papadopoulos2012, GarciaPerez2019, Boguna2021}. In these spaces, nodes have radial (popularity) and angular (similarity) coordinates $r, \theta$, and are connected in the observed network with a probability that decreases with their hyperbolic distance, determined by their coordinates~\cite{Krioukov2010}. Therefore, if the mathematics describing the evolution of the nodes' coordinates was known---and importantly, if this evolution were predictable---one could employ this mathematics to describe and ultimately predict connectivity dynamics. In essence, we observe that network geometry provides a way to cast the problem of network dynamics to a time series prediction problem---a well-mined problem in other areas, such as finance~\cite{Mandelbrot1968, Mandelbrot1997, RLmBm2001, friedrich2011}, network traffic modeling~\cite{ethernetss1993,farimatraffic1999}, and fluid dynamics~\cite{Friedrich1997,Hadjihosseini_2014}.

Given these considerations, here we analyze, for the first time, historical popularity and similarity trajectories of nodes in hyperbolic embeddings of different real networks and find that both types of trajectories exhibit subdiffusive dynamics (see Table~\ref{tab_datasets} for the considered data). Specifically, we find that the trajectories are anti-persistent with short-term negative autocorrelations, and can be well described by a fractional Brownian motion process~\cite{Mandelbrot1968}. 

In hindsight, these findings, and particularly the departure from the traditional law of Brownian motion, agree with intuition. Indeed, real networks are characterized by strong community structures and hierarchical organization that persist over time~\cite{Dorogovtsev10-book}. The first characteristic implies that similarity trajectories should remain confined within specific regions of the similarity space. The second characteristic signifies that popularity trajectories should fluctuate around some expected values that reflect the node positions in the network hierarchy or popularity space. For instance, a non-hub in the US air transportation network is expected to remain a non-hub even though its degree can fluctuate. Further, since the popularity and similarity trajectories are not generally independent, the dynamics of one can influence the dynamics of the other. This picture is analogous to subdiffusive phenomena found in crowded biological systems, where particles diffuse in environments with hierarchies of energy barriers or traps~\cite{Saxton2007}. 

These findings have both practical and theoretical implications. From a practical perspective, the subdiffusivity of the trajectories implies that they are predictable (cf.~Appendix~\ref{sec:pred_examples}), thereby satisfying a necessary condition for predicting connectivity dynamics in real networks. In contrast, as shown, trajectories obtained from synthetic networks that lack an underlying geometry resemble traditional Brownian
motion, and are thus unpredictable. From a theoretical perspective, our findings guide the potential development of dynamical models for the nodes' similarity-popularity motion, akin to Langevin equations in molecular dynamics or Newtonian equations in gravitational mechanics~\cite{brain_theory_2014, schlick2010molecular}. Any such equations for networks, to reflect reality, should give rise to the observed subdiffusive dynamics. 

\begin{table}
\begin{ruledtabular}
\begin{tabular}{lll}
\textbf{Name} &   \textbf{Nodes} & \textbf{No. of snapshots} \\ \hline
US Air &  US airports & 10000 (daily, 1988-2015) \\
Bitcoin &  Bitcoin addresses & 1292 (daily, 2012-2016) \\
PGP WoT & PGP certificates & 1500 (daily, 2003-2007)\\ 
IPv6 Internet & Autonomous sys. & 300 (weekly, 2011-2017) \\ 
arXiv &  Authors & 3977 (daily, 2011-2022) \\
\end{tabular}
\end{ruledtabular}
\caption{
Overview of the considered networks, obtained from Refs.~\cite{USAirlines, btcDownload, pgp_topo_data, as_topo_data_ipv6, arxivDataset} (see Appendix~\ref{sec:data} for details).}
\label{tab_datasets}
\end{table}

More precisely, for each real network, we consider consecutive snapshots of its topology, $G_1, G_2, \ldots, G_\tau$, spanning the period shown in Table~\ref{tab_datasets}. We independently map each snapshot $G_t, t=1, \ldots, \tau$, to an underlying hyperbolic space using Mercator~\cite{GarciaPerez2019} and extract for each node $i$ its popularity and similarity trajectories, i.e., the evolution of its popularity and similarity coordinates, $\{r_i(t), \theta_i(t)\}, t=1, \ldots, \tau$ (Appendix~\ref{sec:embedding}). We choose to independently map the snapshots in order to avoid possible artificial biases between node coordinates across embeddings, and apply Procrustean rotations~\cite{procrustes} to eliminate global rotations and reflections (see Appendix~\ref{sec:procrustes}). 

We note that for each node $i$ its coordinates $r_i(t)$ and $\theta_i(t)$ are inferred from the observed network topology at time $t$, $G_t$, and represent estimates for some underlying or `hidden' true coordinates that determine network connectivity~\cite{Boguna2010}. When $G_t$ changes, these estimates also change. Furthermore, for each node $i$ we  also consider the trajectory of its expected degree $\kappa_i(t)$, which is related to $r_i(t)$ and provides a more direct view on individual node popularity (see Appendix~\ref{sec:embedding}). Figures~\ref{pop_clt}(a),(f) and~\ref{sim_clt}(a) show the popularity and similarity trajectories of the Charlotte Douglas International Airport (CLT) in the US Air transportation network (see Appendix~\ref{pop_sim_examples} for other examples).

\begin{figure}[!t]
\centering
\includegraphics[width=8.5cm]{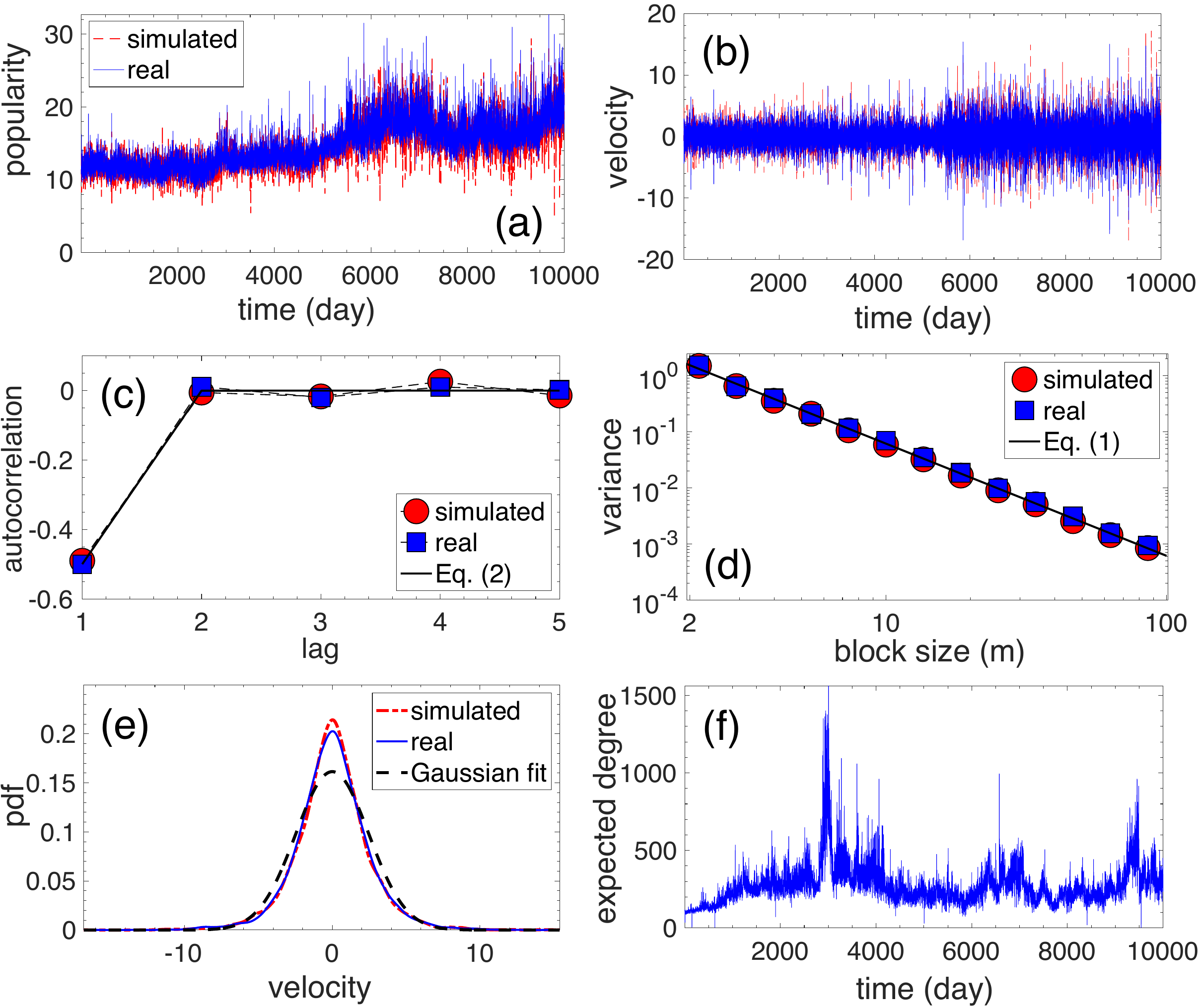}
\caption{Properties of the popularity trajectory of CLT in the US Air (in blue). (a) Radial popularity trajectory. (b) Trajectory increments (velocity). (c) Sample autocorrelation of the velocity. (d) Variance-time plot of the velocity. (e) Probability density function (pdf) of the velocity. The dashed curve shows a Gaussian pdf with the same mean and variance. (f) Expected degree trajectory. The estimated Hurst exponents for the trajectories in (a) and (f) are $0.0004$ and $0.02$. The plots show also results for a simulated counterpart of the trajectory (in red) constructed using the model of Eq.~(\ref{eq:rlfbm_main}).
\label{pop_clt}}
\end{figure}
\begin{figure}[!t]
\centering
\includegraphics[width=8.5cm]{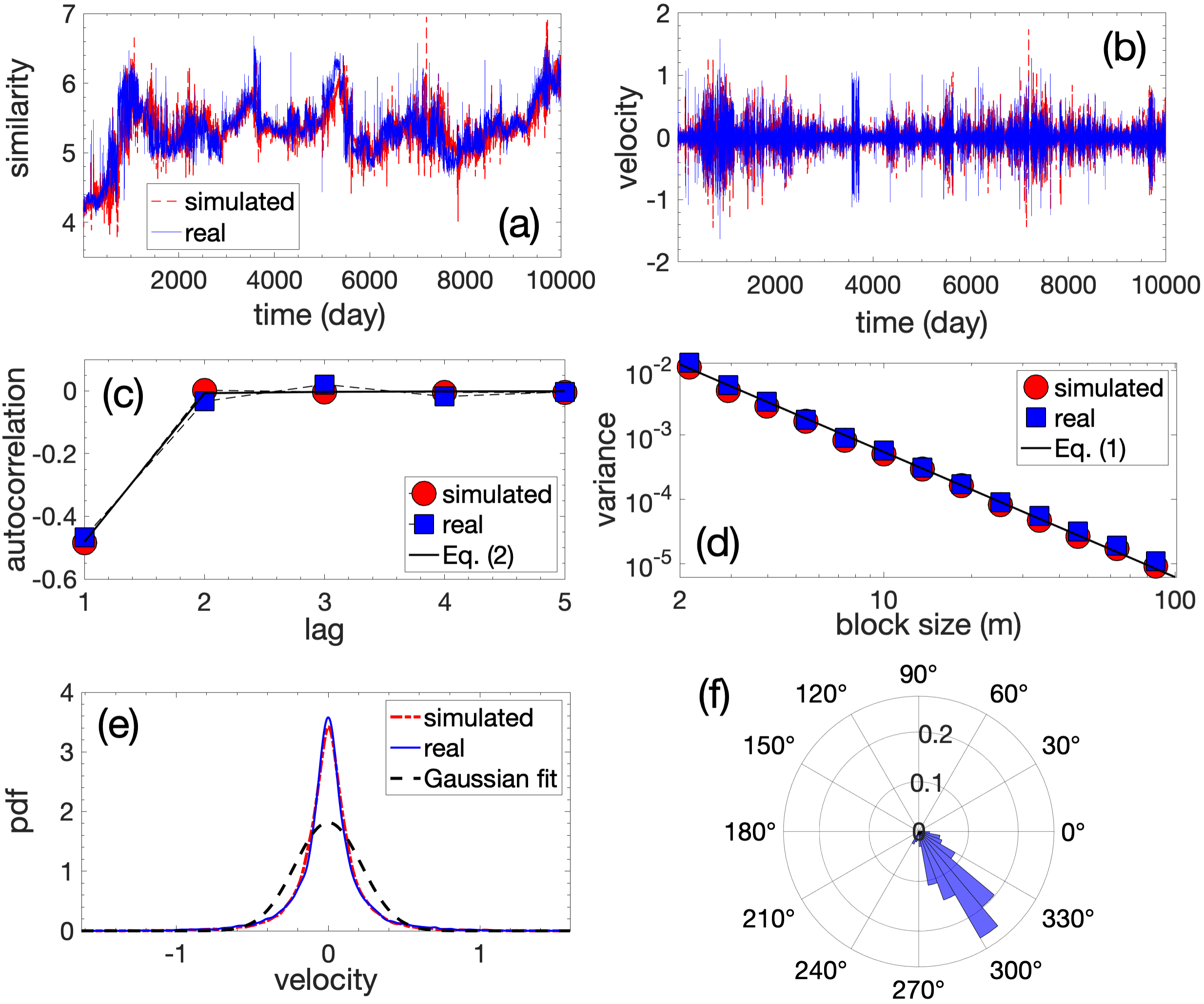}
\caption{Properties of the similarity trajectory of CLT in the US Air (in blue).  (a)-(e) show the same as (a)-(e) in Fig.~\ref{pop_clt}, but for the similarity trajectory. The $y$-axis in~(a) is in radians. (f) Angular distribution of the trajectory. The estimated Hurst exponent for the trajectory is $0.03$.
\label{sim_clt}}
\end{figure}
 
We find that the obtained trajectories constitute well-defined time series, exhibiting universal properties. Particularly, their velocities (or increments, i.e., differences between consecutive observations) manifest negative autocorrelations with short-term memory [see Figs.~\ref{pop_clt}(b),(c), \ref{sim_clt}(b),(c), and Appendix~\ref{sec:unwrapped}]. Furthermore, the velocities exhibit self-similar scale laws of fractional order, i.e., they are of fractal nature [see Figs.~\ref{pop_clt}(d) and~\ref{sim_clt}(d) that are explained below, and Appendix~\ref{pop_sim_examples} for more examples].

A way to quantify the self-similarity and the type of memory inherited by the trajectories is by computing their Hurst exponent $H \in (0,1)$~\cite{Hurst_1951}. 
A value of $H$ in $(0.5,1)$ indicates a persistent time series with long-range positive autocorrelation (a superdiffusive process). On the other hand, a value of $H$ in $(0, 0.5)$ indicates an anti-persistent or `mean-reverting' time series with negative autocorrelation (a subdiffusive process); in this case, positive (negative) increments tend to be followed by negative (positive) increments, and the dependence between increments is short-ranged. The strength of anti-persistence increases as $H$ approaches $0$. The case $H=0.5$ corresponds to a random process with no dependence between its increments (typical diffusion).

\begin{figure}
\centering
\includegraphics[width=8.7cm]{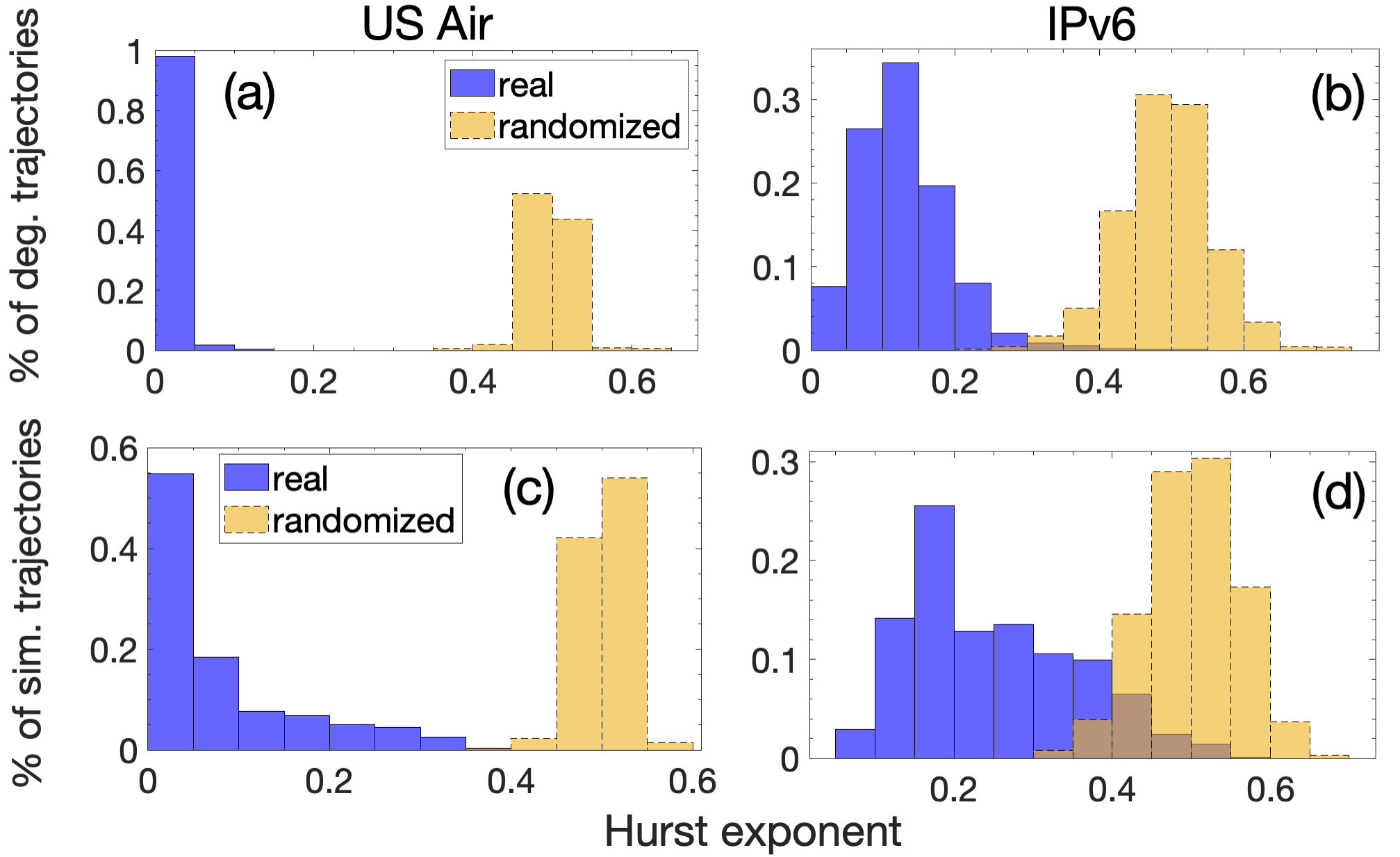}
\caption{Distribution of Hurst exponents in US Air and IPv6 (in blue). (a) and (b) correspond to the expected degree trajectories. The average Hurst exponents are respectively $0.01$ and $0.13$. (c) and (d) correspond to the similarity trajectories. The average Hurst exponents are $0.08$ and $0.25$. We consider trajectories with at least $300$ points. The distributions for the randomized counterparts are shown in yellow. Similar results hold for the radial trajectories (Appendix~\ref{sec:radials}).
\label{hursts}}
\end{figure}

Figure~\ref{hursts} shows the distribution of $H$ across the trajectories in the US Air and IPv6 Internet (similar results hold for the other real networks, see Appendix~\ref{sec:radials}). To calculate $H$ we use the method of absolute moments~\cite{Preis2009, gorski2002} (see Appendix~\ref{sec:hurst_estimation}). We see that in general the distributions are concentrated over values of $H$ well below $0.5$, indicating that the trajectories are generally strongly anti-persistent. The strongest anti-persistence (lowest average $H$) is observed in the US Air and Bitcoin, followed by PGP, IPv6, and arXiv (Fig.~\ref{hursts} and Appendix~\ref{sec:radials}). We note that values of $H$ close to $0$ have also been observed in other real data~\cite{Gatheral2018, Neuman2018}. In general, we find that the popularity trajectories are more anti-persistent than the similarity trajectories. Figure~\ref{hursts} shows also the distributions of $H$ in randomized counterparts of the trajectories, obtained by randomizing the sign of the velocities, which breaks correlations (Appendix~\ref{sec:unwrapped}). As expected, these distributions are concentrated around $H=0.5$.

To further support our findings, we also construct the variance-time plot~\cite{ethernetss1993}. Specifically, let $X_k^{(m)}=(X_{km-m+1}+\ldots+X_{km})/m$, $k=1,2,3,\ldots$, denote an aggregated point series of the velocities over non-overlapping blocks of size $m \geq 2$. Self-similarity implies that the variance of $X^{(m)}$ satisfies
\begin{equation}
\label{eq:var}
\textnormal{Var}(X^{(m)})= \sigma^2 m^{2H-2}, 
\end{equation}
where $\sigma^2$ is the variance of $X$.
We find that the variance-time plots of the velocity processes, i.e., the empirical plots of $\log{\textnormal{Var}(X^{(m)})}$ against $\log{m}$, indeed follow closely Eq.~(\ref{eq:var}) [see Figs.~\ref{pop_clt}(d) and \ref{sim_clt}(d), and Appendix~\ref{pop_sim_examples}]. We also note that qualitatively similar results hold if we consider network snapshots over coarser time intervals such as weekly or monthly intervals (see Appendix~\ref{sec:timescales}).

To understand the origin of the observed anti-persistence, we consider a simple network model where snapshots $G_t$, $t=1,2, \ldots, \tau$, undergo link rewirings (see Appendix~\ref{sec:geom_vs_nongeom}). In the geometric version of the model, the snapshots are constructed according to random hyperbolic graphs (RHGs)~\cite{Krioukov2010}. Here the nodes are assigned fixed similarity coordinates $\theta_h$ and target expected degrees $\kappa_h$, called `hidden variables', and are connected with a probability that decreases with their effective distance, $\Delta\theta_h/(\kappa_h \kappa_h')$, where $\Delta \theta_h$ is the similarity distance and $\kappa_h, \kappa_h'$ are the nodes' expected degrees. In the non-geometric version, the snapshots are constructed either according to the configuration model (CM)~\cite{ChungLu2002} or to random graphs (RGs)~\cite{SoRa51}. In CM the connection probability depends only on the nodes' expected degrees $\kappa_h, \kappa_h'$, which are heterogeneous (Appendix~\ref{sec:geom_vs_nongeom}), while in RGs all nodes have the same expected degree and connected with the same probability. The link-rewiring process involves deleting at random a number of links from snapshot $G_t$ and subsequently reinserting an equal number of links to generate the next snapshot $G_{t+1}$, according to the connection probability in the corresponding model (Appendix~\ref{sec:geom_vs_nongeom}). Once the snapshots are created, we embed them into hyperbolic spaces and extract the nodes' expected degree and similarity trajectories, $\kappa(t)$ and $\theta(t)$, following the same procedure as in the real networks. 

\begin{figure}
\centering
\includegraphics[width=9cm]{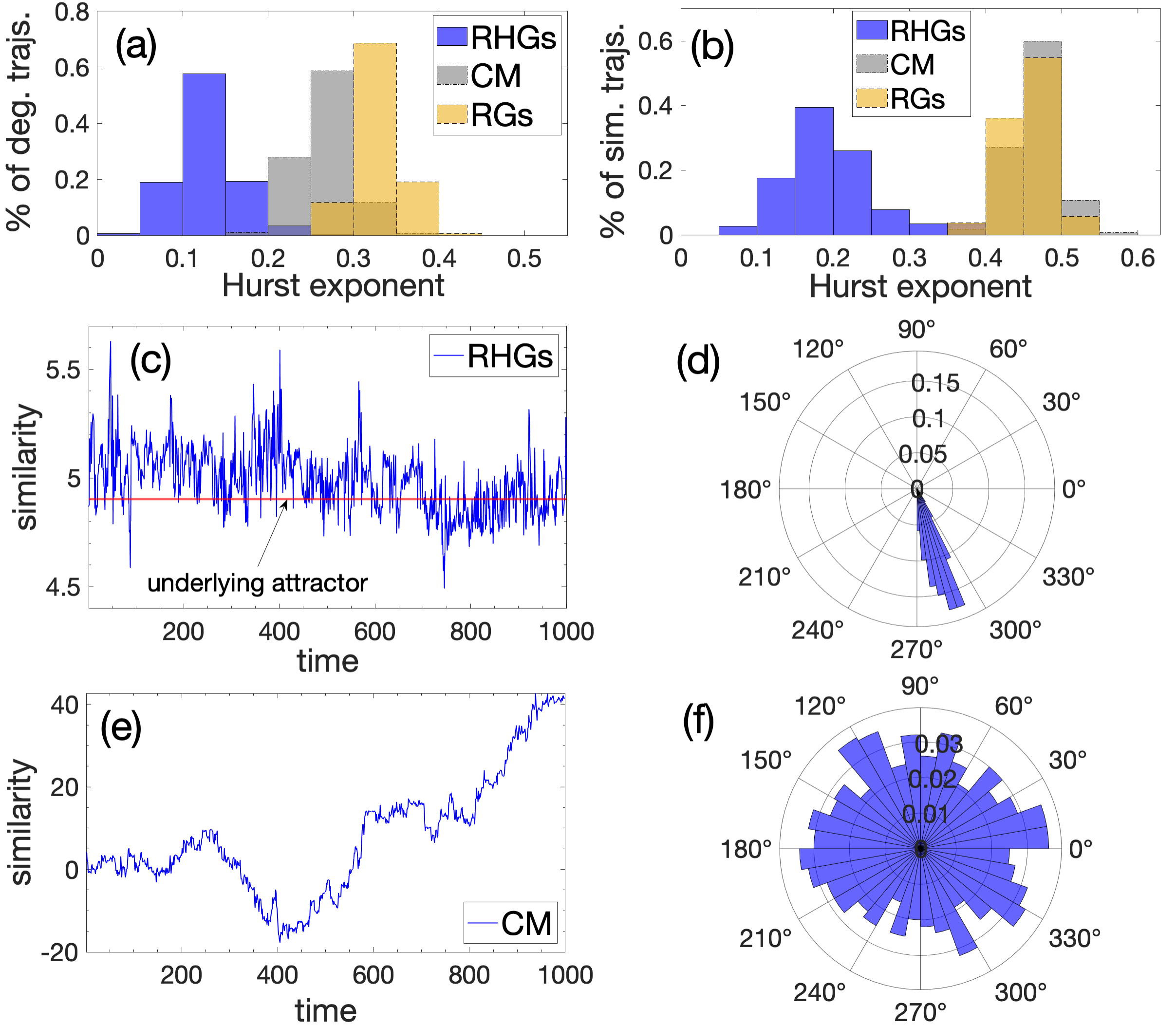}
\caption{Expected degree and similarity trajectories in synthetic temporal networks constructed according to RHGs, the CM, and RGs. (a) and (b) show the distribution of Hurst exponents for the expected degree and similarity trajectories in the three cases. (c) and (d) show a similarity trajectory in RHGs and its distribution in the similarity space. The red line in~(c) indicates the underlying attractor, i.e., the node's hidden similarity coordinate $\theta_h=4.9$. (e) and (f) show the same as in (c) and (d), but for the CM case. There is no underlying attractor here. The Hurst exponents for the trajectories in (c)  and (e) are respectively 0.16 and 0.42.
\label{geom_effect}}
\end{figure}

We find that the trajectories in RHGs exhibit negative autocorrelations and strong anti-persistence, as in real networks [see Fig.~\ref{geom_effect}(a),(b), and Figs.~\ref{sim_s1} and~\ref{deg_s1}]. Further, as in real networks, the similarity trajectories tend to be confined within specific regions of the similarity space [see Fig.~\ref{geom_effect}(c),(d), and Fig.~\ref{sim_clt}(a),(f)]. In contrast, the similarity trajectories in CM and RGs resemble Brownian motion, having Hurst exponents close to $0.5$ and spreading throughout the similarity space  [see Fig.~\ref{geom_effect}(b),(e),(f), and Figs.~\ref{sim_cm} and~\ref{sim_rg}]. Further, the anti-persistence of the degree trajectories weakens [Fig.~\ref{geom_effect}(a) and Figs.~\ref{deg_cm} and~\ref{deg_rg}].

These results can be explained as follows.  In RHGs, even though the snapshots $G_t$ change, they are all created using the same set of underlying node coordinates. Consequently, the inferred coordinates $\kappa(t)$ and $\theta(t)$ for each node should remain close to their respective underlying coordinates $\kappa_h$ and $\theta_h$ at all times $t$. Further, whenever the inferred coordinates happen to deviate significantly from their underlying coordinates, then in the subsequent time step it is more likely that they will be closer to them; otherwise, the inferred coordinates would eventually diverge from their underlying coordinates. Additionally, the inferred coordinates never settle onto their underlying ones, which means that whenever they are sufficiently close to them, then in the subsequent time step they are likely to drift away due to probabilistic network changes. In essence, the underlying coordinates of a node can be seen as `attractors' that pull the inferred coordinates close to them. These attractors are in competition with connectivity changes that push the inferred coordinates away from them. This competition explains the observed reversal nature of the trajectories, i.e., anti-persistence [Fig.~\ref{geom_effect}(c)].

Non-geometric networks lack underlying similarity attractors, and thus $\theta(t)$ changes freely, leading to diffusive dynamics [Fig.~\ref{geom_effect}(e),(f)]. Further, although there are still popularity attractors $\kappa_h$, degree anti-persistence is weaker than in RHGs, as degree trajectories depend on similarity trajectories~\cite{GarciaPerez2019}, which are no longer anti-persistent. When the attractors $\kappa_h$ are heterogeneous, tighter constraints are imposed on individual degrees, supporting the stronger degree anti-persistence in CM compared to RGs in Fig.~\ref{geom_effect}(a).

Taken altogether, these results indicate that the observed anti-persistence of the real-world trajectories is inherently linked to the latent geometry of real networks, i.e., to their intrinsic similarity-popularity space~\cite{Papadopoulos2012}. We note that the underlying popularity and similarity attractors do not have to remain fixed as in the RHGs model, but can change over time, reflecting long-term popularity and similarity shifts. We also note that subdiffusive and self-similar phenomena, explainable through network geometry, have been previously observed in processes running \emph{on} networks and in the characteristics of network structure, cf.~\cite{Boguna2021}. These phenomena are unrelated to our findings here, which are about the dynamics \emph{of} networks. Next, we show that the trajectories can be well described by a fractional Brownian motion process, and explain how their strong anti-persistence favors plausible predictions for their evolution. 

Fractional Brownian motion (fBm) indexed by a Hurst exponent $H$, $B_{H}(t), t \ge 0$, is a popular stochastic process often used as basis in the modeling of real-world time series~\cite{Mandelbrot1968}. The increment process of fBm, $X_H(t)=B_{H}(t+1)-B_{H}(t)$, is called fractional Gaussian noise (fGn) and has the following autocorrelation function~\cite{Mandelbrot1968}:
\begin{equation}
\label{eq:rho}
\rho(l)=0.5\left(|l+1|^{2 H}-2|l|^{2 H}+|l-1|^{2H}\right),
\end{equation}
where $l$ is the time lag. 

Figures~\ref{pop_clt}(c) and~\ref{sim_clt}(c) show that Eq.~(\ref{eq:rho}) closely follows the empirical autocorrelation function of the velocities. However, Figs.~\ref{pop_clt}(e) and~\ref{sim_clt}(e) show that the velocity distribution is not Gaussian, in contrast to fGn (see Appendix~\ref{pop_sim_examples} for other examples). Further, the velocities are characterized by time-varying variances [cf.~Figs.~\ref{pop_clt}(b) and~\ref{sim_clt}(b), and Appendix~\ref{pop_sim_examples}]. These facts suggest that standard fBm is rather too simplistic to fully capture the trajectory characteristics. Instead, we find below that an fBm with time-varying noise-induced variance adequately captures the trajectories.

Specifically, we consider a modification of the Riemann-Liouville multifractional Brownian motion model of Ref.~\cite{RLmBm2001}, where instead of varying the Hurst exponent over time, we vary the noise-induced variance (see Appendix~\ref{sec:fbm_model}, and Appendix~\ref{sec:multifractality} for discussion on multifractality). The model aims to capture trajectories that behave only locally, i.e., within intervals of variance stationarity, as an fBm. Further, we note that trajectories may exhibit both trends and a mean-reverting behavior in their increments. Therefore, we also adjust the model to account for possible trends in the trajectories that can also change with time. The final equation for the considered fBm model takes the form
\begin{align}
\label{eq:rlfbm_main}
\tilde{B}_H(t)&=B_0+\int_{0}^{t}\mu(s)\mathrm{d}s\\
\nonumber &+\frac{1}{\Gamma{(H+\frac{1}{2})}}\int_{0}^{t}(t-s)^{H-1/2}\sigma(s)\mathrm{d}B(s),~~t \geq 0,
\end{align}
where $B(s)$ is the standard Brownian motion, $B_0$ is the initial position, $H \in (0,1)$ is the Hurst exponent, $\Gamma$ is the gamma function, and $\mu(s)$ and $\sigma(s)$ are respectively the trend and noise-induced volatility at time $s$. In Appendix~\ref{sec:fbm_model}, we provide the discrete-time analogue of the model and explain how to tune its parameters to create simulated counterparts of real trajectories.

Figures~\ref{pop_clt}(a) and~\ref{sim_clt}(a) show that the model can adequately capture the popularity and similarity trajectories of CLT in the US Air. In general, we find that the model can capture the trajectories of all the considered networks (see Figs.~\ref{USair_ex}-\ref{arXiv_kappa} for other examples), apart from some cases (present mainly in PGP and arXiv) where similarity trajectories appear to exhibit jumps. Such cases suggest that extensions that also account for jumps may be desirable as part of future work~\cite{Xiao2010}. 

Finally, given the ability of the considered fBm model to simulate trajectories resembling real ones, a natural next question is whether the model can also be used for predicting the future evolution of the trajectories. While a comprehensive assessment of the model's predictive capabilities is beyond the scope of the present work, we provide a glimpse on the model's ability for predictions in Appendix~\ref{sec:pred_examples}. We show that predictions are possible, even with simple educated guesses on the model's parameters for the prediction period, based on historical data. This predictability can be explained by the fact that the variance of the considered model grows with time $t$ as $t^{2H}$~\cite{Mandelbrot1968}. Therefore, as real trajectories are typically characterized by low values of $H \ll 0.5$, their variance grows slowly with time, remaining even approximately constant for cases where $H \approx 0$, favoring their predictability. 

Our results pave the way towards the ultimate goal of predicting connectivity dynamics in real networks over different time scales. To accomplish this goal, robust methodologies for trajectory prediction should be developed through automated approaches for fine-tuning the parameters of the proposed model or of possible variations of it. Of particular interest are techniques that can predict changes in the expected trends of the trajectories. In this vein, an interesting direction involves analyzing many trajectories simultaneously, instead of individually, which could identify possible synchronization phenomena across the trajectories and guide the development of dynamical models. Predicting connectivity dynamics would then be equivalent to predicting the evolution of hyperbolic distances among node pairs and deciding on their temporal connectivity based on their distance.
\newline 
\newline
Datasets and code used in the paper are available at~\cite{code_and_data}.  
\begin{acknowledgments}
The authors acknowledge support by the TV-HGGs project (OPPORTUNITY/0916/ERC-CoG/0003), co-funded by the European Regional Development Fund and the Republic of Cyprus through the Research and Innovation Foundation. 
\end{acknowledgments}

\appendix
\section{Real-world network data}
\label{sec:data}

Here we provide details on the considered real-world networks. An overview of the data is given in Table~\ref{tab_datasets} of the main text. 

\begin{figure*}
\centering
\includegraphics[width=17.5cm]{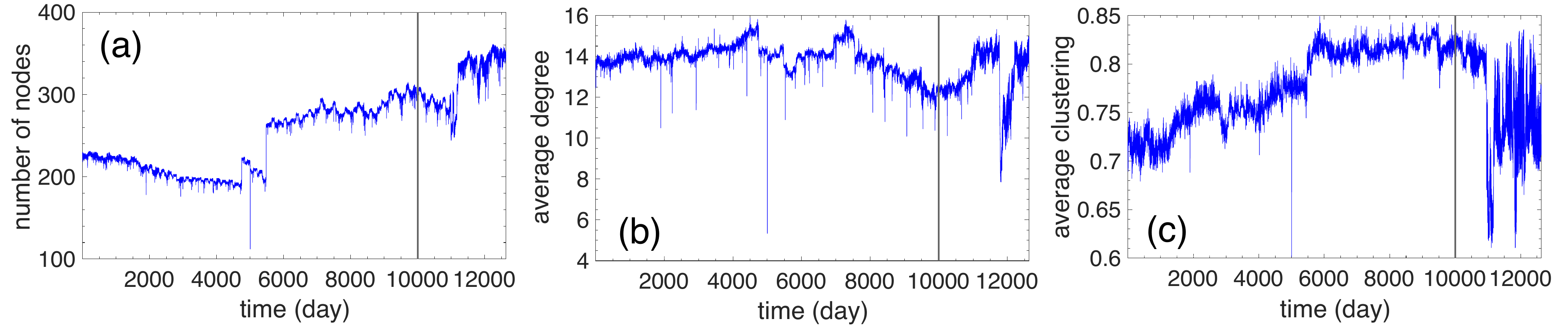}
\vspace{-0.3cm}
\caption{(a) Number of nodes, (b) average degree, and (c) average clustering in the US Air. We consider data up to time 10000 (May 20, 2015) indicated by the vertical line in the plots.
\label{USAir_props}}
~\\
\includegraphics[width=17.5cm]{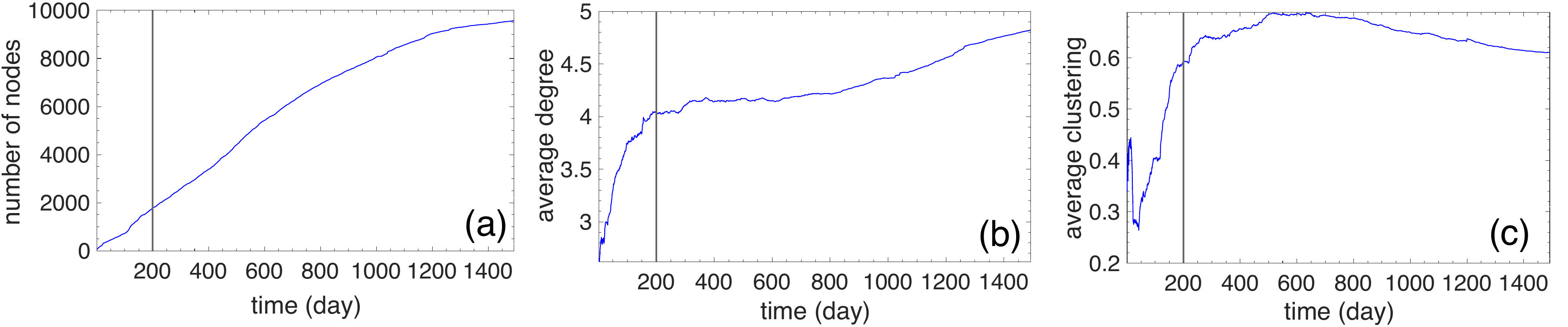}
\vspace{-0.3cm}
\caption{(a) Number of nodes, (b) average degree, and (c) average clustering in the BTC. We consider data after time 200 (August 22, 2012).
\label{BTC_props}}
~\\
\includegraphics[width=17.5cm]{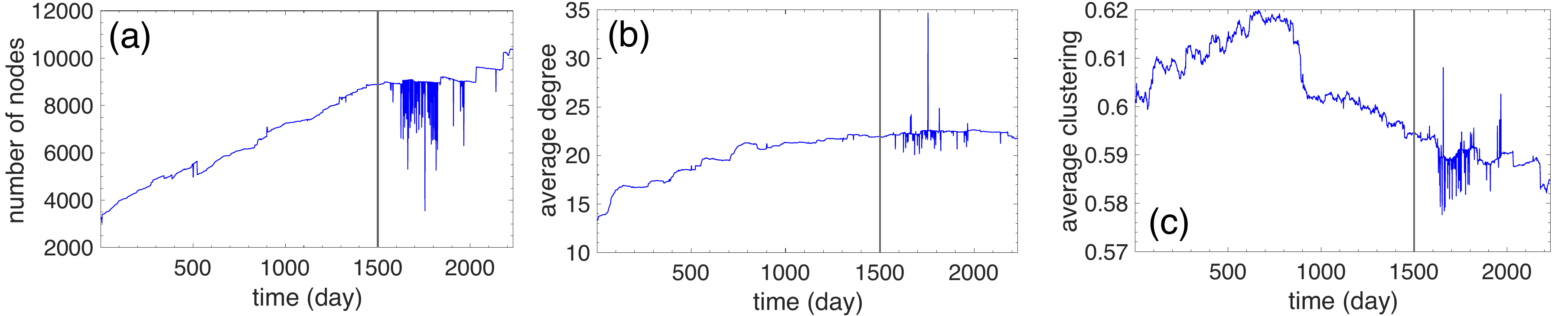}
\vspace{-0.3cm}
\caption{(a) Number of nodes, (b) average degree, and (c) average clustering in the PGP WoT. We consider data up to time 1500 (May 4, 2007).
\label{PGP_props}}
~\\
\includegraphics[width=17.5cm]{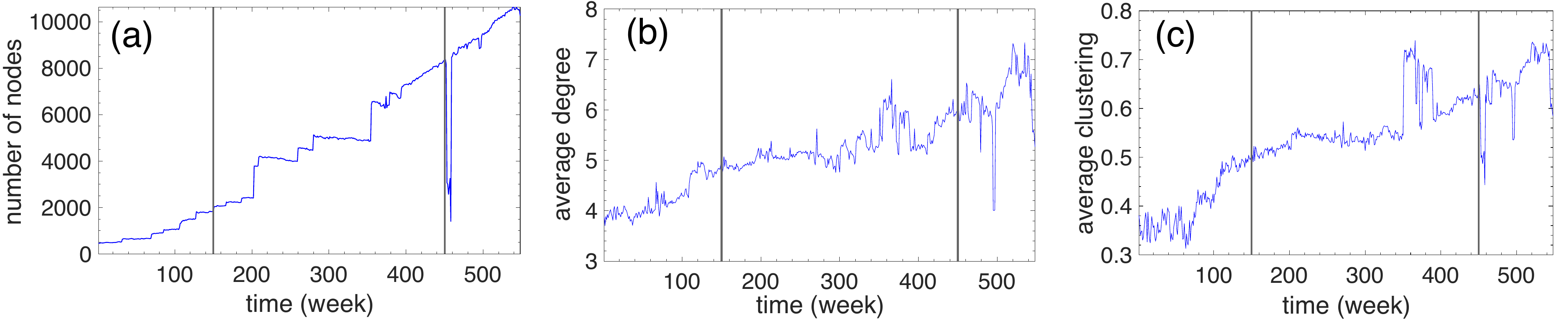}
\vspace{-0.3cm}
\caption{(a) Number of nodes, (b) average degree, and (c) average clustering in the IPv6 Internet. We consider data from time 150 to 450 (October 17, 2011 to July 17, 2017).
\label{IPv6_props}}
~\\
\includegraphics[width=17.5cm]{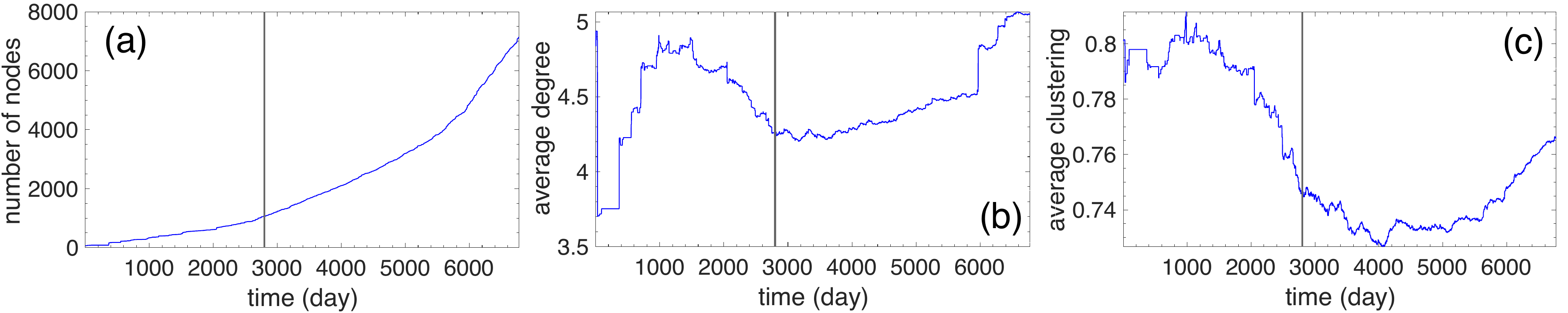}
\vspace{-0.3cm}
\caption{(a) Number of nodes, (b) average degree, and (c) average clustering in the arXiv (Quantitative Finance). We consider data after time 2800 (August 1, 2011).
\label{arXiv_props}}
\end{figure*}

\textbf{US Air.} The US Air network corresponds to domestic flights between US airports. The data set is constructed by data available by the US department of transportation~\cite{USAirlines}. Each directed link signifies a flight between a source and destination airport in the US. We consider 10000 topology snapshots corresponding to the period January 1988 to May 2015. Each snapshot is obtained by merging the flight links between the US airports over a one day period. For each of the snapshots we form their undirected counterparts by taking into account only bi-directional links between airports. For each snapshot of the undirected counterparts we isolate the largest connected component. The evolution of the number of nodes, average degree, and average clustering~\cite{Dorogovtsev10-book} in the network is shown in Fig.~\ref{USAir_props}. In the considered period these quantities evolve in a relatively stable manner. We do not consider data after the aforementioned period as the average clustering in daily snapshots fluctuates rapidly (Fig.~\ref{USAir_props}(c)). 

We note that in the considered period, the percentage of new links between consecutive snapshots, i.e., the percentage of links present in snapshot $G_t$ but not in $G_{t-1}$ is on average $1.6\%$ with a standard deviation of $1.8$. For the old links between consecutive snapshots, i.e., the links present in snapshot $G_{t-1}$ but not in $G_{t}$, these numbers are respectively $1.6\%$ and $1.9$.

\textbf{Bitcoin (BTC).} The BTC network corresponds to transactions between bitcoin addresses in the bitcoin cryptocurrency network and is constructed by data taken from Ref.~\cite{btcDownload}. Each directed link in the network signifies a bitcoin transaction between a sender and a receiver address. We consider 1292 daily topology snapshots corresponding to the period August 2012 to March 2016. Each snapshot is obtained by merging all the transaction links up to the date of the snapshot. For each of the snapshots we form their undirected counterparts by taking into account only bi-directional transaction links between bitcoin addresses. For each of the undirected counterparts we consider links attached to nodes with degrees $k \geq 4$, and isolate the largest connected component. The evolution of the number of nodes, average degree, and average clustering in the network is shown in Fig.~\ref{BTC_props}. In the considered period these quantities evolve in a relatively stable manner. The percentage of new links between consecutive snapshots is on average $0.14\%$ with a standard deviation of $0.14$. The network is purely growing and therefore there are no link removals.

\textbf{PGP WoT.} Pretty-Good-Privacy (PGP)  is an encryption program that provides cryptographic privacy and authentication for data communication~\cite{openpgp}. PGP web of trust (WoT) is a directed network where nodes are certificates consisting of public PGP keys and owner information. A directed link in the web of trust pointing from certificate {\it A} to certificate {\it B} represents a digital signature by owner of {\it A} endorsing the owner/public key association of {\it B}. We use temporal PGP web of trust data collected by J\"{o}rgen
Cederl\"{o}f~\cite{pgp_topo_data}. Specifically, we consider $1500$ daily PGP topology snapshots spanning the period March 2003 to May 2007. For each of the snapshots we form their undirected counterparts by taking into account only bi-directional trust links between the certificates. For each of the undirected counterparts we consider links attached to nodes with degrees $k \geq 6$, and isolate the largest connected component. The evolution of the number of nodes, average degree, and average clustering in the network is shown in Fig.~\ref{PGP_props}. We do not consider data after the aforementioned period as the number of nodes fluctuates rapidly (Fig.~\ref{PGP_props}(a)). 
The percentage of new links between consecutive snapshots is on average $0.16\%$ with a standard deviation of $0.52$. For the old links between consecutive snapshots these numbers are respectively $0.06\%$ and $0.49$.
 
\textbf{IPv6 Internet.} The IPv6 Autonomous Systems (AS) Internet topology snapshots were extracted from data collected by CAIDA~\cite{ark2009}. Pairs of ASs peer to exchange traffic and the links in the AS topology represent peering relationships between ASs. CAIDA's IPv6 data set~\cite{as_topo_data_ipv6} provides regular snapshots of AS links. The data set consists of ASs that can route packets with IPv6 destination addresses. We consider 300 topology snapshots, spanning the period October 2011 to July 2017. Each snapshot corresponds to an interval of one week and is obtained by merging the AS links observed in that week. The evolution of the number of nodes, average degree, and average clustering in the network is shown in Fig.~\ref{IPv6_props}. In the considered period these quantities evolve in a relatively stable manner. 
The percentage of new links between consecutive snapshots is on average $6.3\%$ with a standard deviation of $3.8$. For the old links between consecutive snapshots these numbers are respectively $5.8\%$ and $2.3$.

\textbf{arXiv.} The temporal arXiv collaboration network is constructed by data taken from Ref.~\cite{arxivDataset}. In arXiv, each paper is assigned to one or more relevant categories. We consider the temporal co-authorship network formed by the authors of papers in the category ``Quantitative Finance". In each topology snapshot the nodes are authors that are connected if they have co-authored a paper. We consider 3977 daily snapshots corresponding to the period August 2011 to July 2022. Each snapshot is obtained by merging all the co-author links up to the date of the snapshot, and isolating the largest connected component. The evolution of the number of nodes, average degree, and average clustering in the network is shown in Fig.~\ref{arXiv_props}. The percentage of new links between consecutive snapshots is on average $0.05\%$ with a standard deviation of $0.17$. The network is purely growing and therefore there are no link removals.

\section{Hyperbolic embedding}
\label{sec:embedding}

We map each network snapshot to an underlying hyperbolic space using Mercator~\cite{GarciaPerez2019}. Mercator combines the Laplacian Eigenmaps approach of Ref.~\cite{carlo1} with maximum likelihood estimation, used also in previous methods~\cite{Boguna2010,frag:hypermap, frag:hypermap_cn}, to produce fast and accurate embeddings. In a nutshell, Mercator takes as input the network's adjacency matrix $A$. The generic element of the matrix is $A_{ij}=A_{ji}=1$ if there is a link between nodes $i$ and $j$, and $A_{ij}=A_{ji}=0$ otherwise. It then infers radial (popularity) and angular (similarity) coordinates, $r_i$ and $\theta_i$, for all nodes $i \leq N$. To this end, it maximizes the likelihood function
\begin{equation}
\label{eq:likelihood}
\mathcal L=\prod_{1 \leq j < i \leq N} p(x_{ij})^{A_{ij}}\left[1-p(x_{ij})\right]^{1-A_{ij}},
\end{equation}
where the product goes over all node pairs $i, j$ in the network, while $p(x_{ij})$ is the Fermi-Dirac connection probability,
\begin{equation}
\label{eq:px}
p(x_{ij})=\frac{1}{1+e^{\frac{1}{2T}(x_{ij}-R)}}.
\end{equation}
Here, $x_{ij}=r_i+r_j+2 \ln{(\Delta\theta_{ij}/2)}$ is approximately the hyperbolic distance between nodes $i$ and $j$~\cite{Krioukov2010}; $\Delta\theta_{ij}=\pi-|\pi-|\theta_i-\theta_j||$ is the angular similarity distance; $R$ is the radius of the hyperbolic disc where nodes reside; and $T \in (0,1)$ is the network temperature, which is also inferred by Mercator and is related to the clustering strength of the network. 

The connections and disconnections among nodes act respectively as attractive and repulsive forces. Mercator feels these attractive/repulsive forces, placing connected (disconnected) nodes closer to (farther from) each other in the hyperbolic space. We note that changes in the adjacency matrix $A$ result in the re-evaluation of all node positions.

The radial coordinate $r$ of a node is related to its expected degree $\kappa$, as
\begin{equation}
\label{r_k}
r \sim R-2\ln{\kappa}. 
\end{equation}
The disc radius $R$ grows logarithmically with the network size $N$, $R \sim 2\ln{N}$, and depends also on the average degree and clustering in the network~\cite{Krioukov2010, GarciaPerez2019}. The evolution of $R$ in the considered networks is shown in Fig.~\ref{hyperbolic_discs}. 

Furthermore, as mentioned in the main text, we also consider the trajectories of expected degrees, i.e., the trajectories of $\kappa$ in~(\ref{r_k}).\footnote{To be precise, $\kappa$ in~(\ref{r_k}) is proportional to the node's expected degree in the network; it is exactly equal to it only for a uniform distribution of angular coordinates and at $N \to \infty$~\cite{GarciaPerez2019, Krioukov2010}. Nevertheless, without loss of generality, we call $\kappa$ expected degree.} We consider these, as they do not directly depend on the evolution of $R$, and are thus more direct proxies to the evolution of individual node popularities. In contrast, $R$ can significantly dominate in~(\ref{r_k}), rendering the evolution of individual popularities via radial coordinates less apparent.

The code implementing Mercator is made publicly available at~\cite{mercatorcode}. We have used the code without any modifications. To minimize fluctuations due to random number generation we use the same random seed value for embedding the snapshots of a given network.
 
\section{Procrustean rotations} 
\label{sec:procrustes}

The distances between the nodes in the embeddings are invariant with respect to global rotations and reflections of the angular coordinates. Thus, given the embeddings $E_{t-1}$ and $E_t$ of two consecutive snapshots $G_{t-1}$ and $G_t$, the angles in $E_t$ can be globally shifted compared to the angles in $E_{t-1}$. To mitigate such effects, we apply the following procedure. 

Let $\mathcal{S}=E_1, E_2, \ldots, E_\tau$ be the sequence of embeddings of snapshots $G_1, G_2, \ldots, G_\tau$. We consider the nodes' angular trajectories in the sequence of optimally rotated embeddings, $\mathcal{S^{\textnormal{r}}}=E_1^{\textnormal{r}}, E_2^{\textnormal{r}}, \ldots, E_\tau^{\textnormal{r}}$, obtained as follows. First, $E_1^{\textnormal{r}}$ is the same as $E_1$. Then, each successive embedding $E_t^{\textnormal{r}}$, $t=2, \ldots, \tau$, is obtained by globally shifting the nodes' angles in $E_t$ such that the sum of the squared distances (SSD) between the nodes' angles in $E_t$ and the angles of the corresponding nodes in $E_{t-1}^{\textnormal{r}}$ is minimized. To this end, we apply a Procrustean rotation~\cite{procrustes}, as follows:

\begin{enumerate}
\item Let $\{\theta_{t-1}^i\}$ and $\{\theta_t^i\}$ be the sets of node angles in $E_{t-1}^{\textnormal{r}}$ and $E_t$, respectively. We transform the angles to Cartesian coordinates $\{x_i, y_i\}=\{\cos{\theta_{t-1}^i}, \sin{\theta_{t-1}^i}\}$ and $\{w_i, z_i\}=\{\cos{\theta_t^i},\sin{\theta_t^i}\}$. 
\item A rotation of the points $\{w_i, z_i\}$ by an angle $\phi$ is given by
\begin{align}
\nonumber \{u_i, v_i\}=\{w_i \cos \phi-z_i \sin \phi,\\
w_i \sin\phi + z_i \cos \phi\},
\end{align}
where $u_i, v_i$ are the coordinates of the rotated point $w_i, z_i$. The SSD between $\{u_i, v_i\}$ and $\{x_i, y_i\}$ is 
\begin{equation}
\textnormal{SSD}=\sum_{i \in \mathcal{C}} (u_i-x_i)^2+ (v_i-y_i)^2.
\end{equation}
The sum is taken over the set of nodes $\mathcal{C}$ that exist in both $E_{t-1}^{\textnormal{r}}$ and $E_t$. The optimal rotation angle $\phi^{*}$ is computed by taking the derivative of the SSD with respect to $\phi$ and solving for $\phi$ when the derivative is zero,
\begin{equation}
\phi^{*}=\tan^{-1}\left[\frac{\sum_{i \in \mathcal{C}} (w_i y_i - z_i x_i)}{\sum_{i \in \mathcal{C}}(w_i x_i + z_i y_i)} \right].
\end{equation}
The optimally rotated angles are then computed as 
\begin{equation}
\{\theta_{\textnormal{rotated}}^i\}=\{\tan^{-1}(v_i^*/u_i^*)\},
\end{equation}
where  
\begin{align}
\nonumber \{u_i^*, v_i^*\}=\{w_i \cos{\phi^*}-z_i \sin{\phi^*},\\
 w_i \sin{\phi^*} + z_i\cos{\phi^*}\}.
\end{align}
We note that if $\theta_{\textnormal{rotated}}^i < 0$, then $\theta_{\textnormal{rotated}}^i \coloneqq 2 \pi+\theta_{\textnormal{rotated}}^i$.
\item We repeat the above procedure after replacing $\{\theta_t^i\}$ with $\{2\pi-\theta_t^i\}$, which is the reflection of the former across the $x$-axis, and compute the optimally rotated angles in this case as well, $\{\tilde{\theta}_{\textnormal{rotated}}^i\}$. 
\item Finally, we compute $D_\theta=\sum_{i \in \mathcal{C}}|\theta_{\textnormal{rotated}}^i-\theta_{t-1}^i|$ and $\widetilde{D}_\theta=\sum_{i \in \mathcal{C}}|\tilde{\theta}_{\textnormal{rotated}}^i-\theta_{t-1}^i|$. The optimally rotated angles are  $\{\theta_{\textnormal{rotated}}^i\}$ if $D_\theta < \widetilde{D}_\theta$, and $\{\tilde{\theta}_{\textnormal{rotated}}^i\}$ otherwise. $E_t^{\textnormal{r}}$ is obtained from $E_t$ by replacing the nodes' angles in the latter with their optimally rotated angles.
\end{enumerate}

\section{Hurst exponent estimation}
\label{sec:hurst_estimation}

There exists a variety of techniques for estimating the Hurst exponent $H$ of a time series~\cite{taqqu1995}. Here we use the method of absolute moments or generalized exponents~\cite{Preis2009, gorski2002}, which involves calculating the sample moments of a time series at different lags and then fitting a linear regression model to estimate $H$.  

Specifically, for a time series $g(t)$ with $t \in \{1, 2, \ldots, \tau\}$ the time lag-dependent generalized Hurst exponent, $H_q(l)$, can be determined by the general relationship~\cite{Preis2009, gorski2002} 
\begin{equation}
\label{eq:sql}
S_q(l)=\langle|g(t+l)-g(t)|^q \rangle \propto l^{q H_q(l)},
\end{equation}
where $q > 0$, while $l \ll \tau$ is the time lag, $l = 1, 2, 3, \ldots$. The brackets $\langle \ldots \rangle$ denote the expectation value over $t$.

To estimate the Hurst exponent $H$ of a time series we leverage Eq.~(\ref{eq:sql}). First, we compute $S_q(l)$ for different lags $l=1,2, \ldots, l_\textnormal{max}$, and orders $q=1, 2, \ldots, q_{\textnormal{max}}$. Then, for each order $q$, we find the best fit line (in a least-squares sense) for $\log{S_q(l)}$ against $\log{l}$. From the slope $\lambda$ of this line we estimate the generalized Hurst exponent $H_q=\lambda/q$. Finally, we estimate $H$ as the average of $H_q$. In our analysis, we use $q_{\textnormal{max}}=3$ and $l_\textnormal{max}=10$, except for the arXiv where we use $l_\textnormal{max}=20$ (as $l_\textnormal{max}=10$ in this case appeared to sometimes overestimate $H$). 

We note that for the radial trajectories $r(t)$, $g(t)$ in Eq.~(\ref{eq:sql}) is simply $g(t)=r(t)$. The same holds for the degree trajectories $\kappa(t)$. For the similarity trajectories $\theta(t)$, the situation is a bit more involved as the motion takes place on a circle. In that case, Eq.~(\ref{eq:sql}) is not applied on $\theta(t)$ per se, but on the unwrapped similarity trajectories $\theta_{\textnormal{u}}(t)$, i.e., $g(t)=\theta_{\textnormal{u}}(t)$, where $\theta_{\textnormal{u}}(t)$ is computed as described in Appendix~\ref{sec:unwrapped}. 

We also considered other methods for estimating $H$, such as rescaled range (R/S) analysis~\cite{taqqu1995}, reaching similar conclusions. We prefer the method of moments described above, as we found it the most accurate in estimating low values of $H$. See also Appendix~\ref{sec:multifractality}, where we discuss multifractality.

\section{Velocities and unwrapped similarity trajectories}
\label{sec:unwrapped}

The increment or velocity process $v(t), t=2,3, \ldots$, for the radial trajectories is simply given by $v(t)=r(t)-r(t-1)$, and for the degree trajectories by $v(t)=\kappa(t)-\kappa(t-1)$. For the similarity trajectories it is given by
\begin{equation}
\label{eq:vtheta}
v(t)=s\Delta\theta(t), 
\end{equation}
where $\Delta\theta(t)=\pi-|\pi-|\theta(t)-\theta(t-1)||$ is the angular distance between positions $\theta(t)$ and $\theta(t-1)$, and $s$ is $1$ or $-1$ depending on the direction of the motion (counterclockwise or clockwise). For all trajectories we define $v(1)=0$. 

The direction $s$ in Eq.~(\ref{eq:vtheta}) can be computed using the cross product. Specifically, let
\begin{align}
\{u_x, u_y\}=&\{\cos{\theta(t-1)},~\sin{\theta(t-1)}\},\\
\{v_x, v_y\}=&\{\cos{\theta(t)},~\sin{\theta(t)}\}.
\end{align}
Then,
\begin{equation}
s=\text{sgn}(u_x v_y-u_y v_x),
\end{equation}
where  $\text{sgn}(x)$ is the sign function.

The unwrapped similarity trajectories $\theta_\text{u}(t)$, are given by 
\begin{equation}
\theta_\text{u}(t)=\theta(1)+\sum_{i=1}^{t} v(t).
\end{equation}
Intuitively, $\theta_\text{u}(t)$ is an unfolding of $\theta(t)$, such that the similarity motion takes place on the real line. This allows us to directly compute quantities, like the one given by Eq.~(\ref{eq:sql}), which assume that the process is evolving on an unbounded domain. In contrast, in the original representation, $\theta(t)$, the motion takes place on the circle $[0, 2\pi]$, i.e., on a bounded domain with periodic boundary conditions (a node passing through one side of the domain $[0, 2\pi]$, appears on the other side). 

\textbf{Plotting similarity trajectories.} Freed by boundaries, the representation $\theta_\text{u}(t)$ also allows us to more clearly visualize the similarity motion. We use this representation whenever we plot similarity trajectories (i.e., we plot $\theta_\text{u}(t)$ instead of $\theta(t)$). We note that one can revert back to $\theta(t)$ from $\theta_\text{u}(t)$ as follows. Let $m(t)=\textnormal{mod}[\theta_\text{u}(t), 2\pi]$. Then, $\theta(t)=m(t)$ if $m(t) > 0$, otherwise $\theta(t)=2\pi+m(t)$. 

\textbf{Randomized trajectories.} Let $v(t), t = 2,3, \ldots$, be the velocity process of a real trajectory, and $X(1)$ the trajectory's initial value. To construct the randomized counterpart of the trajectory, $X^{\textnormal{r}}(t)$, we proceed as follows. First, we randomize the signs of the velocities, while preserving their magnitude, i.e., we create the velocity process $v^{\textnormal{r}}(t)=\tilde{s}|v(t)|$, where $\tilde{s}=1$ or $\tilde{s}=-1$ with probability $0.5$. Then, we compute $X^{\textnormal{r}}(t)$ as
\begin{equation}
X^{\textnormal{r}}(t)=X(1)+\sum_{i=1}^{t}v^{\textnormal{r}}(t).
\end{equation}

\section{Evolution of hyperbolic discs and Hurst exponent distributions}
\label{sec:radials}

\begin{figure*}[!t]
\includegraphics[width=16cm]{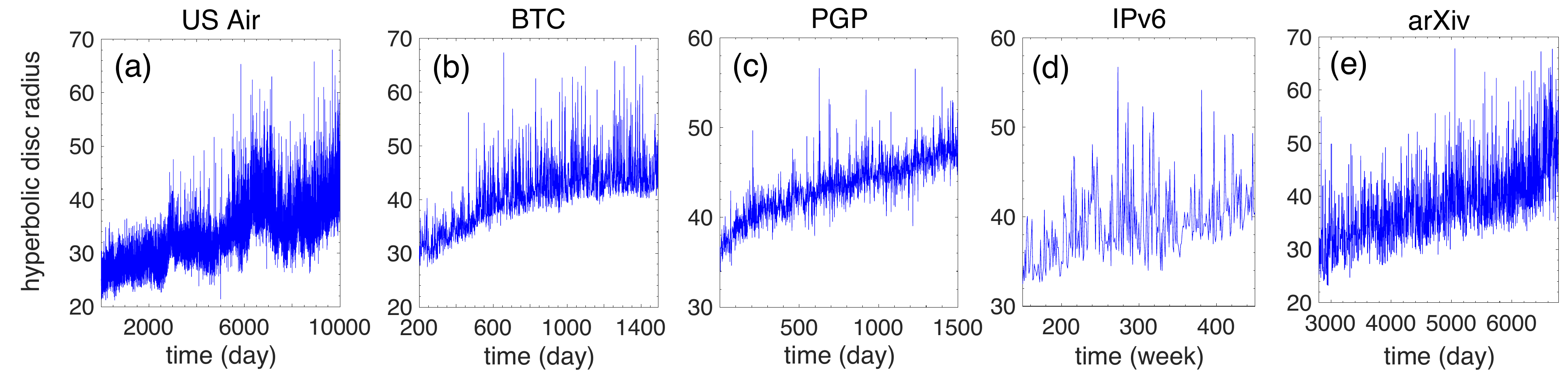}
\caption{Evolution of the hyperbolic disc radius in the considered networks. The estimated Hurst exponents in (a)-(e) are respectively $0.0008$, $0.003$, $0.007$, $0.003$, and $0.11$. 
\label{hyperbolic_discs}}
\end{figure*}

\begin{figure*}[!t]
\centering
\includegraphics[width=14cm]{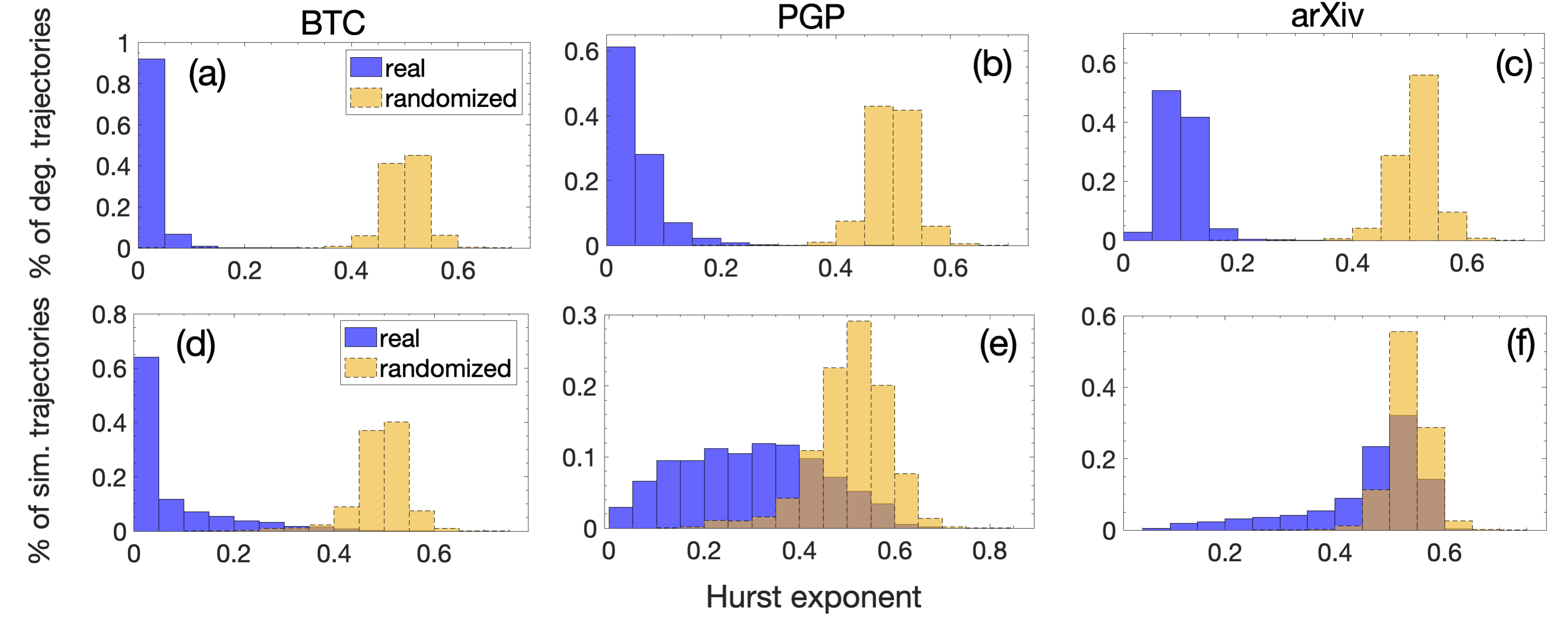}
\caption{Same as in Fig.~\ref{hursts} of the main text, but for BTC, PGP, and arXiv. The average Hurst exponents for the degree trajectories in (a)-(c) are respectively $0.02$, $0.05$, and $0.1$. For the similarity trajectories in (d)-(f) they are $0.07$, $0.30$, and $0.46$.
\label{hursts_rest}}
\end{figure*}

\begin{figure*}[!t]
\centering
\includegraphics[width=16.5cm]{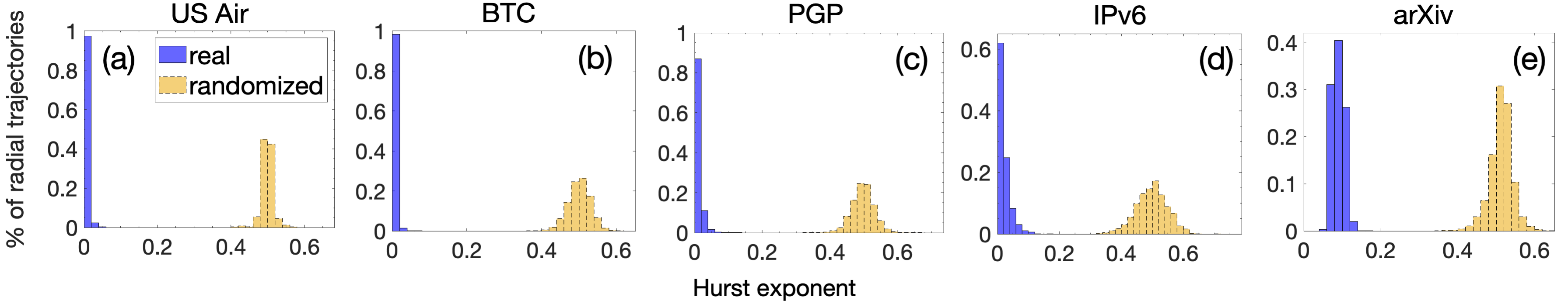}
\caption{Distributions of Hurst exponents for the radial trajectories in the considered real networks (in blue) and for the randomized counterparts of the trajectories (in yellow). The average Hurst exponents in (a)-(e) for the real trajectories are respectively $0.004$, $0.005$, $0.01$, $0.02$, and $0.09$.}
\label{radial_hursts}
\end{figure*}

Figure~\ref{hyperbolic_discs} shows the evolution of the radius $R$ of the hyperbolic disc, where nodes reside, in the considered networks. We see that in all cases the trajectory of $R$ is also strongly anti-persistent. Further, Fig.~\ref{hursts_rest} shows the distributions of Hurst exponents for the expected degree and similarity trajectories in BTC, PGP, and arXiv. Finally, Fig.~\ref{radial_hursts} shows the distributions of Hurst exponents for the radial popularity trajectories in all considered networks.

\section{Fractional Brownian motion model}
\label{sec:fbm_model}

As mentioned in the main text, the trajectories can be adequately captured by a fractional Brownian motion model. Specifically, we build on the fractional Brownian motion based on the Riemann-Liouville fractional integral (RL-fBm), defined as
\begin{equation}
\label{eq:rlfbm}
B_H(t)=\frac{1}{\Gamma{(H+\frac{1}{2})}}\int_{0}^{t}(t-s)^{H-1/2}\mathrm{d}B(s),~~t \geq 0,
\end{equation}
where $B(s)$ is the standard Brownian motion, $H \in (0,1)$ is the Hurst exponent, and $\Gamma$ is the gamma function, cf.~\cite{RLmBm2001}. The RL-fBm corresponds to processes with finite starting time, and for sufficiently large times $t$ it is equivalent to the standard fBm~\cite{Lim2001}. 

To generate $N$ points from an RL-fBm, one can use the discrete simulation scheme described in Refs.~\cite{RLmBm2001, Rambaldi1994}. In that scheme, Eq.~(\ref{eq:rlfbm}) is approximated by
\begin{equation}
\label{eq:rlfbmdis}
B_H(t_j)=\frac{1}{\Gamma{(H+\frac{1}{2})}}\sum_{i=1}^{j} \int_{(i-1)\Delta t}^{i\Delta t} (t_j-\tau)^{H-1/2}\mathrm{d}B(\tau),
\end{equation}
where $\Delta t=1/(N-1)$ is the time step; $t_j=j \Delta t$, $j=1,2, \ldots, N$, are the discrete time points; and $\mathrm{d}B(\tau)$ is the increment of Brownian motion,
\begin{equation}
\mathrm{d}B(\tau)=\left(\frac{\xi_i}{\sqrt{\Delta t}}\right)d\tau,
\end{equation}
where $\xi_i$ is a discrete sequence of Gaussian white noise with zero mean and unit variance. Upon integration Eq.~(\ref{eq:rlfbmdis}) gives
\begin{equation}
\label{eq:sim1}
B_H(t_j)= \sum_{i=1}^j \xi_i w_{j-i+1}\sqrt{\Delta t} ,
\end{equation}
where $w_i$ is a weighting function, whose improved form is~\cite{RLmBm2001, Rambaldi1994}
\begin{equation}
\label{eq:sim2}
w_i=\frac{1}{\Gamma{(H+\frac{1}{2})}}\left[\frac{t_i^{2H}-(t_i-\Delta t)^{2H}}{2 H \Delta t}\right]^{1/2}.
\end{equation}
Equations~(\ref{eq:sim1}) and~(\ref{eq:sim2}) form the basis of the simulation scheme. Code implementing this scheme is available at~\cite{rlfbmsim}.

To simulate trajectories we use a generalization of the above scheme, which accounts for a possibly changing trend and variance, as well as for a non-zero initial position. In this generalization, $B_H(t_j)$ in Eq.~(\ref{eq:sim1}) is replaced by
\begin{equation}
\label{eq:oursim}
\tilde{B}_H(t_j)=B_0+\sum_{i=1}^j \left(\mu_i+\sigma_i \xi_i w_{j-i+1}\sqrt{\Delta t} \right).
\end{equation}
In the last relation, $B_0$ is the initial position, while $\mu_i$ and $\sigma_i$ are respectively the trend and noise-induced volatility at time step $t_i$. Specifically, in our case $B_0$ is the starting position of the popularity or similarity trajectory we want to simulate, $H$ is the trajectory's estimated Hurst exponent, and $\mu_i$ and $\sigma_i$ are estimated from the data as described below. The continuous-time analogue of Eq.~(\ref{eq:oursim}) is Eq.~(\ref{eq:rlfbm_main}) in the main text. 
\subsection{Estimation of $\mu_i$}
\label{sec:mu_est}

Given the velocity series $v(t)$, $t = 1, 2, \ldots$, of a popularity or similarity trajectory (computed as described in Appendix~\ref{sec:unwrapped}), we estimate $\mu_i$, $i=1,2,\ldots$, using the following exponentially weighted moving average scheme
\begin{align}
\nonumber \mu_1&=v(1),\\
\mu_i&=\alpha v(i)+(1-\alpha)\mu_{i-1},~~~i=2,3,\ldots.
\end{align}
Here, $\alpha \in [0,1]$ denotes the smoothing factor. Lower values of $\alpha$ assign less weight to recent measurements and short-term fluctuations, allowing for capturing longer-term trends. For all of our computations we use $\alpha=0.01$, except for the angular trajectories of the IPv6, where we use $\alpha=0.1$ (here a higher value of $\alpha$ appeared necessary in order to better capture the trajectories).

\subsection{Estimation of $\sigma_i$}
\label{sec:sigma_est}

Let $v(t)$, $t = 1, 2, \ldots, \tau$, be the velocity series of a popularity or similarity trajectory (Appendix~\ref{sec:unwrapped}). For each point $t$ we consider a window of length at most $k$ around it, i.e., we consider all $t' \in [\text{max}\{1, t-k/2\}, \text{min}\{t+k/2, \tau\}]$, and compute the local growth rate  
\begin{equation}
\label{eq:sqk}
S_k(t)=\langle|v(t')|\rangle_k.
\end{equation}
The brackets $\langle \ldots \rangle_k$ denote the expectation value over $t' \in [\text{max}\{1, t-k/2\}, \text{min}\{t+k/2, \tau\}]$, while $k/2$ is assumed to be an integer. 

Now, consider the increments $v(t)=B_H(t+\Delta t)-B_H(t)$ of a discretized fBm process $B_H(t)$, where $\Delta t$ is the time step. It can be shown, cf.~\cite{bianchi2005}, that
\begin{equation}
\label{eq:gfbm}
\langle|v(t)|\rangle= \frac{\sqrt{2}}{\Gamma{(\frac{1}{2})}}\sigma_H \sigma \Delta t^{H},
\end{equation}
where $\sigma$ is the noise-induced volatility, while $\sigma_H$ is the $H$-induced volatility,
\begin{equation}
\sigma_H=\sqrt{\frac{\Gamma{(1-2H)}\cos{(H \pi)}}{H \pi}}.
\end{equation}
For the case of $\tilde{B}_H(t_j)$ in Eq.~(\ref{eq:oursim}), we can assume that Eq.~(\ref{eq:gfbm}) holds locally, i.e., within a stationarity window of length $k$ around each point $t_j$. Therefore, to have similar local growth rates as in the real trajectory, we estimate $\sigma_i$, $i = 1, 2, \dots$, as
\begin{equation}
\sigma_i=\frac{\Gamma{(\frac{1}{2})}}{\sqrt{2} \sigma_H}S_k(i) \Delta t^{-H},
\end{equation}
where $S_k(i)$ is given by Eq.~(\ref{eq:sqk}) and $\Delta t=1/(N-1)$. 

The above approach for computing local noise-induced volatilities is similar in spirit to the approach used in Ref.~\cite{RLmBm2001} for computing local Hurst exponents. As in Ref.~\cite{RLmBm2001}, for all of our computations we arbitrarily fix $k=8$, bearing in mind that smaller values of $k$ give better accuracy but larger fluctuations, and vice versa.

\section{Multifractality considerations}
\label{sec:multifractality}

We also considered the possibility of multifractality, i.e., the idea that the trajectories may not be characterized by a single value of $H$, but by a time-varying $H$~\cite{RLmBm2001}. To this end, we examined the singularity spectrum of the trajectories~\cite{wendt2007, jaffard2007}, and investigated if more sophisticated modeling is required in which the trajectories are modeled using time-varying Hurst exponents~\cite{RLmBm2001}. These investigations did not yield substantial evidence that multifractality is present in the trajectories or that a more sophisticated multifractal description is required. Specifically, our investigations suggested that the trajectories can be regarded as monofractal (characterized by a single value of $H$) with a time-varying noise-induced variance. This motivated the development of the fBm model of the previous section. And indeed, we see that this model can capture the real trajectories and their properties very well (cf. Figs.~\ref{pop_clt},~\ref{sim_clt} in the main text and Figs.~\ref{USair_ex}-\ref{arXiv_kappa} in the next section). 

We note that the estimation of $H$ in Appendix~\ref{sec:hurst_estimation} as an average of the estimated generalized exponents $H_q$, $q \in [1, q_{\textnormal{max}}]$, makes sense only under the assumption of monofractality, where $H_q=H,~\forall q$. In our analysis, we use $q_{\textnormal{max}}=3$, as our trajectories are finite and non-stationary (exhibit time-varying variance). Under these conditions, the estimation of $H_q$ becomes less reliable as $q$ increases, as it depends on higher order moments that are more affected by big fluctuations and generally require more samples to be reliably estimated.

We also note that in stationary time series that exhibit multifractality, $H_q$ decreases monotonically with $q$~\cite{barabasi1991, Jiang_2019, schumann2011, ludescher2011}. However, this simple test alone cannot be used to reliably determine if our trajectories exhibit multifractality, since as mentioned above, the trajectories exhibit non-stationarities and have finite lengths. Under such conditions, even purely monofractal data can appear multifractal, while a decreasing estimated $H_q$ does not necessarily imply multifractality. In fact, in such cases the estimated $H_q$ may have a non-monotonic behavior and can even increase with $q$, behaviors that we have observed in our trajectories, and which have been documented in prior studies~\cite{schumann2011, ludescher2011}. Nevertheless, it is still worth checking the $H_q$~vs.~$q$ plots to see if they can generally support the conclusion that the trajectories do not possess multifractality.

\begin{figure*}
\centering
\includegraphics[width=16cm]{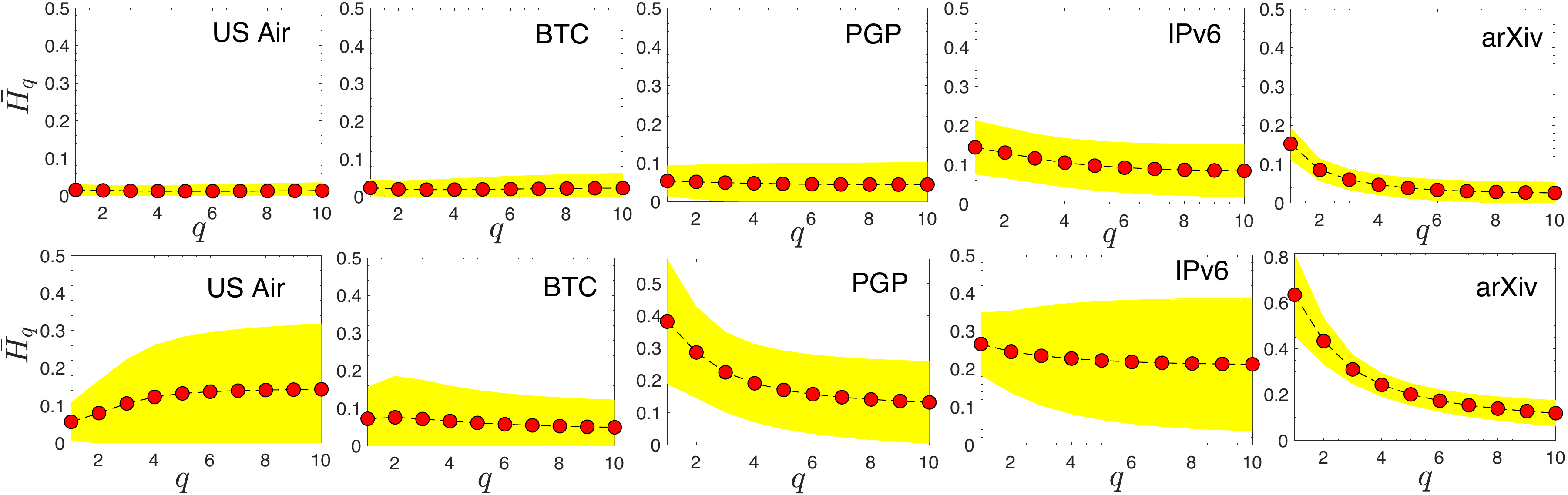}
\caption{Estimated Hurst exponent $H_q$~vs.~$q$ for the expected degree trajectories (top row) and for the angular similarity trajectories (bottom row) in the considered networks. In each case the results represent averages across all trajectories. Shaded areas stand for one standard deviation away from the average. Results for the radial popularity trajectories are similar to the ones in the first row.
\label{fig:H_q}}
\end{figure*}

Results are shown in Fig.~\ref{fig:H_q}. The results indicate that the trajectories can be generally regarded as monofractal, corroborating our previous conclusion. For the angular trajectories of PGP and arXiv, $H_q$ decreases in a more pronounced manner. However, this behavior is also observed in the synthetic counterparts of the trajectories, which are constructed using our monofractal fBm model. Therefore, while we cannot totally exclude the possibility of multifractality, the observed behavior in these cases is likely due to the inaccurate estimation of $H_q$ as $q$ increases, as these trajectories often exhibit larger fluctuations and even jumps. The results also support averaging $H_q$ up to some $q_{\textnormal{max}}$ to obtain a reasonable approximation for the overall scaling behavior, while the cutoff $q_{\textnormal{max}}$ could be probably optimized depending on the case. 

\section{Examples of popularity-similarity trajectories}
\label{pop_sim_examples}

\begin{figure*}
\centering
\includegraphics[width=17.5cm]{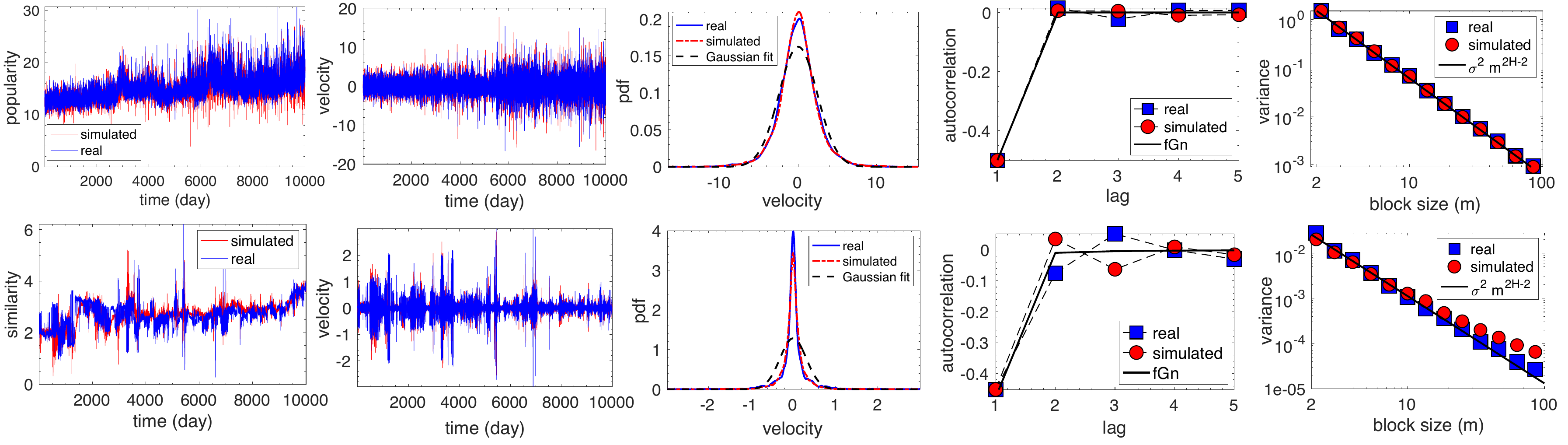}
\vspace{-0.3cm}
\caption{US Air, node LAX (Los Angeles International Airport). $H_\textnormal{pop}=0.001$, $H_\textnormal{sim}=0.03$.}
\label{USair_ex}
~\\
\includegraphics[width=17.5cm]{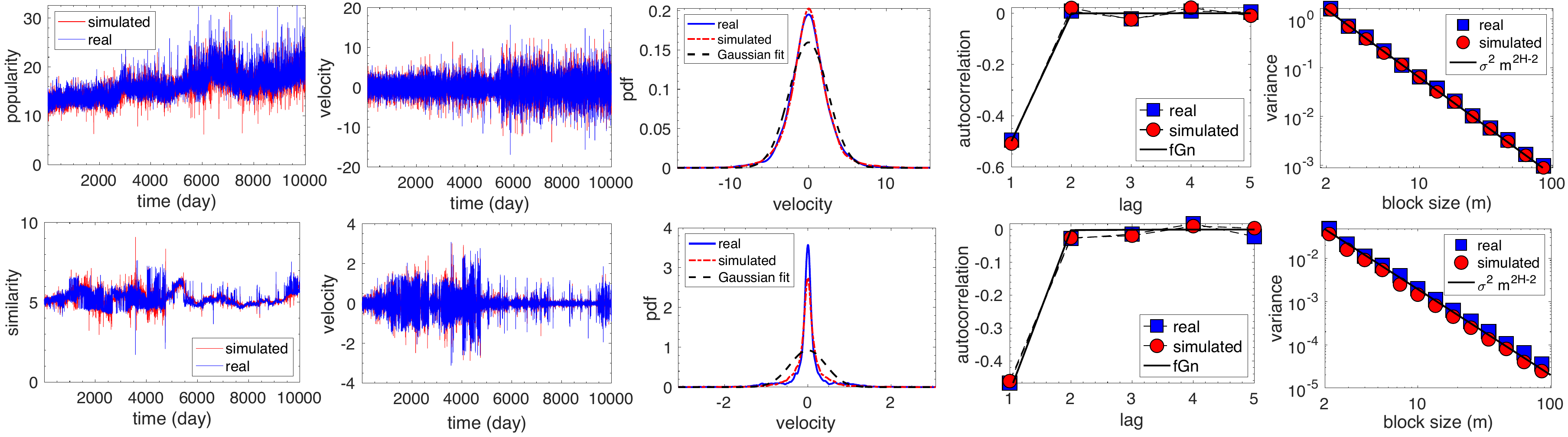}
\vspace{-0.3cm}
\caption{US Air, node BOS (Boston Logan International Airport). $H_\textnormal{pop}=0.001$, $H_\textnormal{sim}=0.01$.}
~\\
\includegraphics[width=17.5cm]{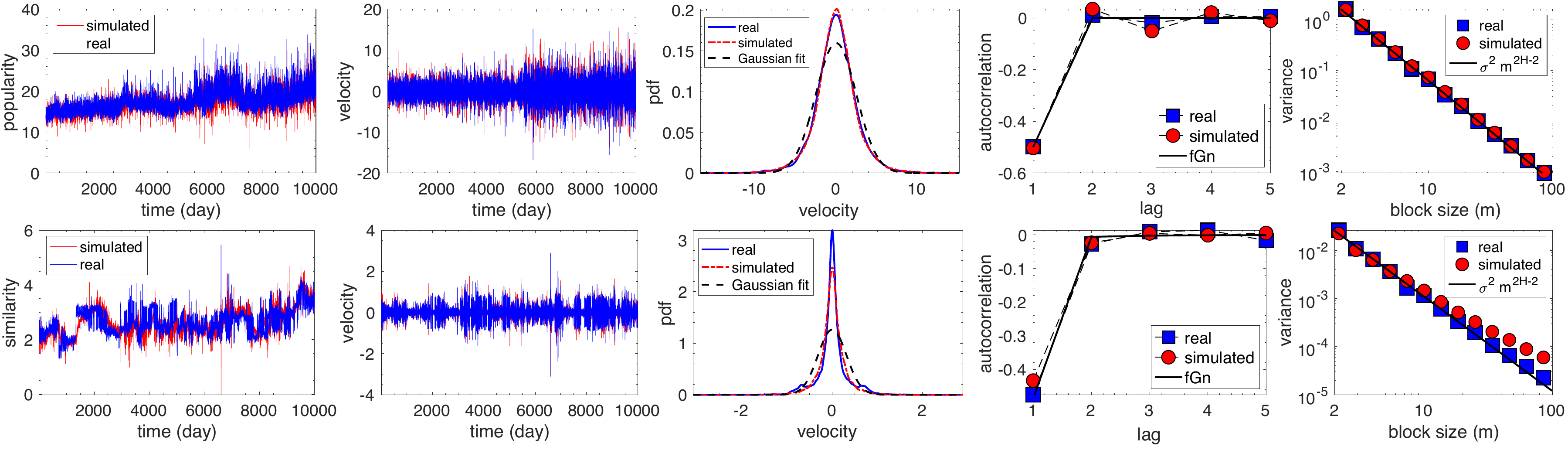}
\vspace{-0.3cm}
\caption{US Air, node SAN (San Diego International Airport). $H_\textnormal{pop}=0.001$, $H_\textnormal{sim}=0.02$.}
~\\
\includegraphics[width=17.5cm]{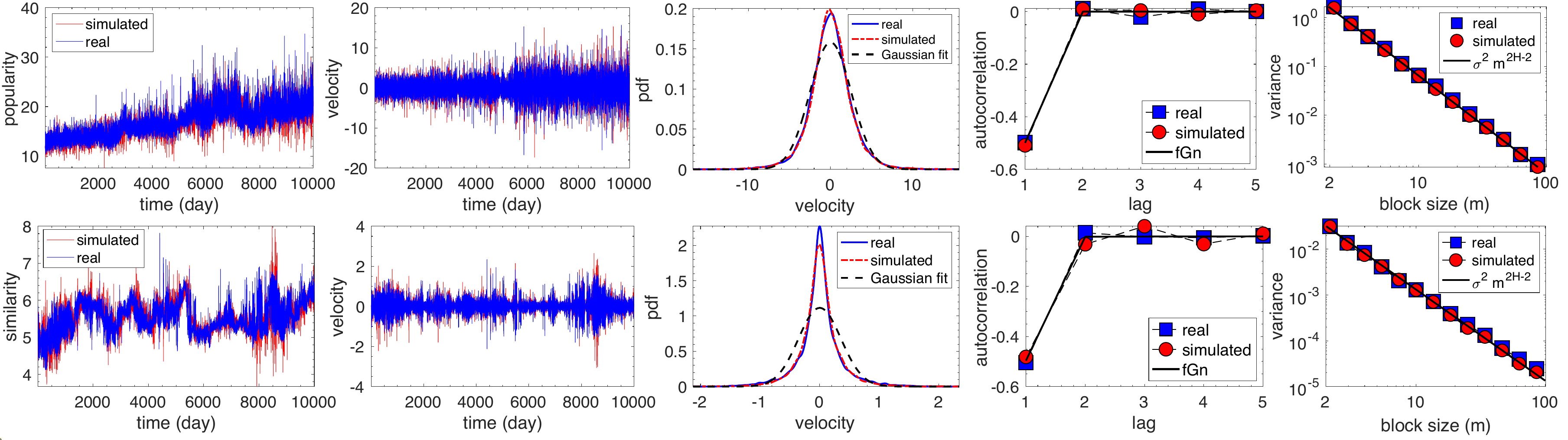}
\vspace{-0.3cm}
\caption{US Air, node LGA (LaGuardia Airport). $H_\textnormal{pop}=0.001$, $H_\textnormal{sim}=0.005$.}
\end{figure*}

\begin{figure*}
\centering
\includegraphics[width=17.5cm]{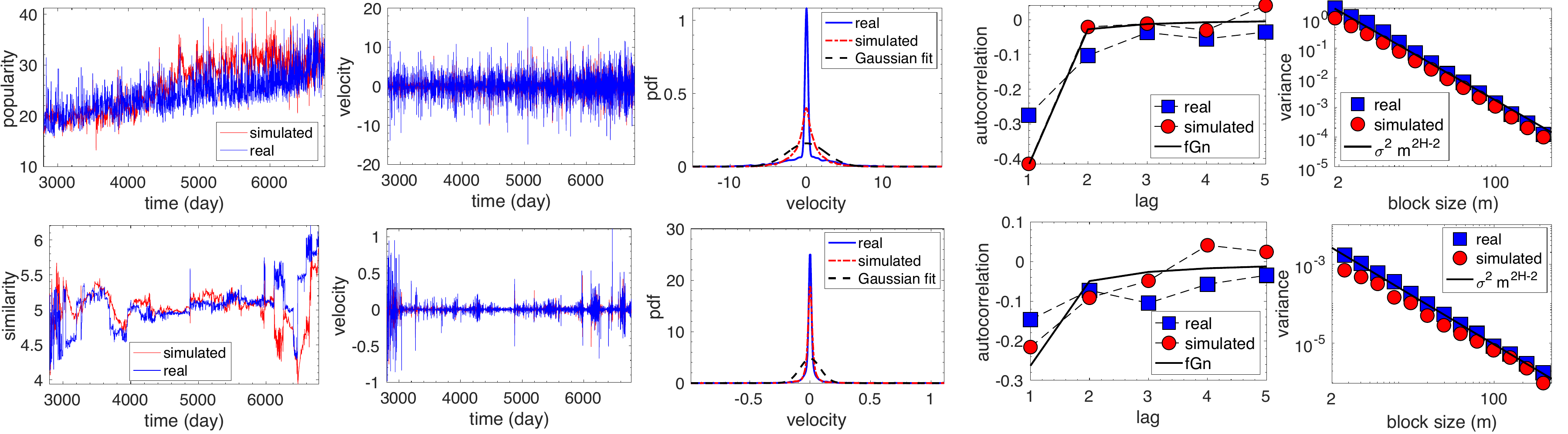}
\vspace{-0.3cm}
\caption{arXiv, node 1058. $H_\textnormal{pop}=0.11$, $H_\textnormal{sim}=0.28$.}
~\\
\includegraphics[width=17.5cm]{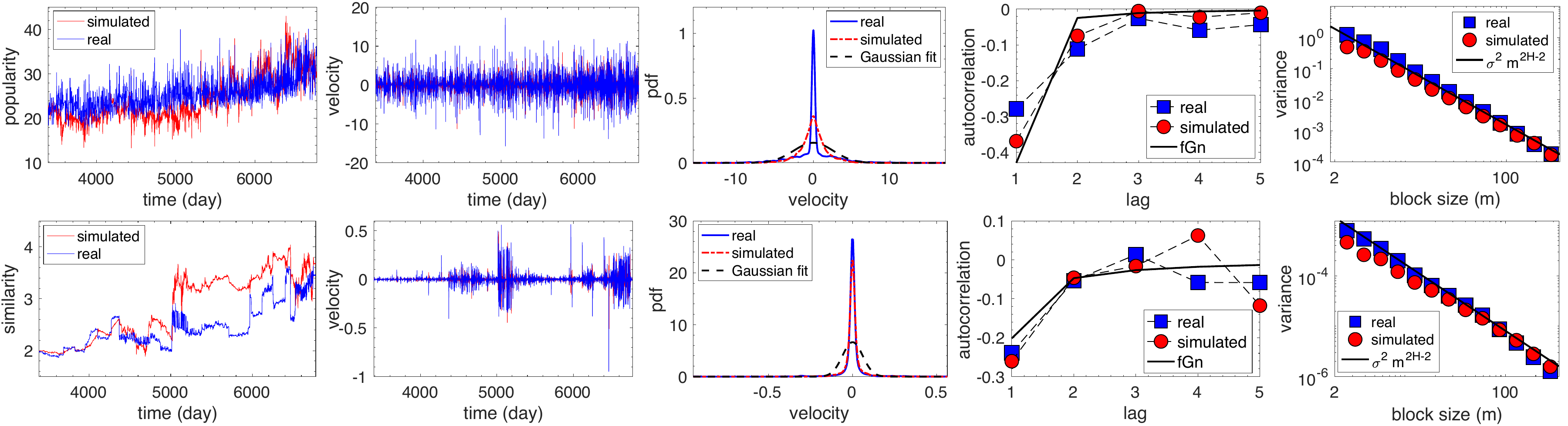}
\vspace{-0.3cm}
\caption{arXiv, node 1218. $H_\textnormal{pop}=0.10$, $H_\textnormal{sim}=0.34$.}
~\\
\includegraphics[width=17.5cm]{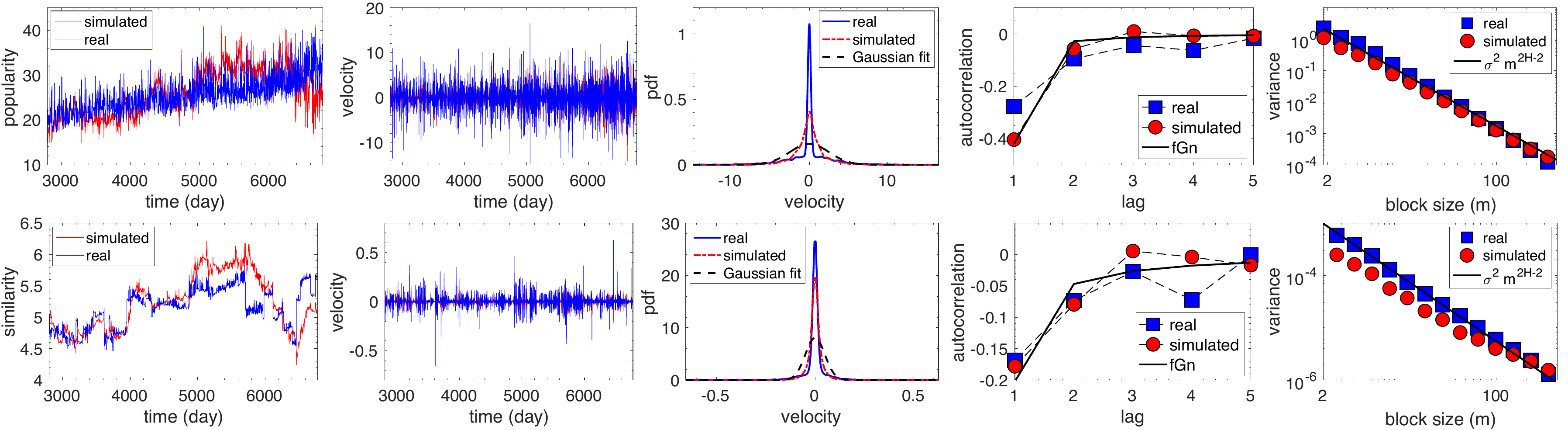}
\vspace{-0.3cm}
\caption{arXiv, node 1827. $H_\textnormal{pop}=0.11$,  $H_\textnormal{sim}=0.34$.}
~\\
\includegraphics[width=17.5cm]{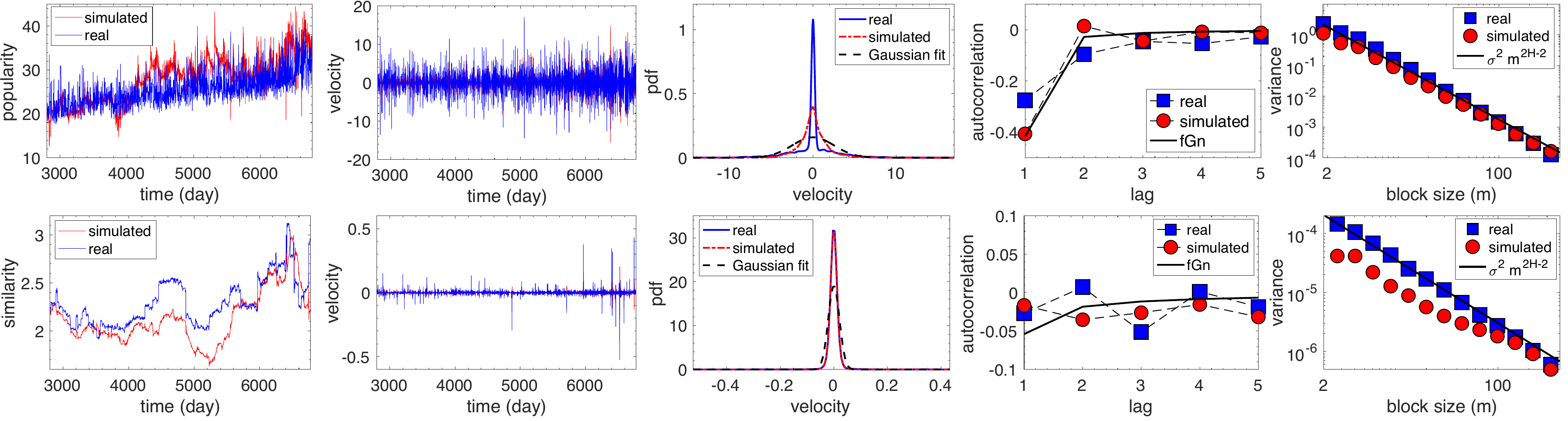}
\vspace{-0.3cm}
\caption{arXiv, node 1674. $H_\textnormal{pop}=0.11$, $H_\textnormal{sim}=0.46$.}
\label{arXiv_ex}
\end{figure*}

Figures~\ref{USair_ex}-\ref{arXiv_ex} show examples of radial popularity and angular similarity trajectories in the considered real networks, as in Figs.~\ref{pop_clt} and~\ref{sim_clt} of the main text. As in the main text, the figures also show realizations of simulated counterpart trajectories (in red color) constructed as described in Appendix~\ref{sec:fbm_model}. The caption in each figure indicates the network, the id of the corresponding node in the data, and the estimated Hurst exponents of its popularity and similarity trajectories, denoted respectively by $H_\textnormal{pop}$ and $H_\textnormal{sim}$, and computed as described in Appendix~\ref{sec:hurst_estimation}. Figures~\ref{USAir_kappa}-\ref{arXiv_kappa} show examples of expected degree trajectories. In each case the estimated Hurst exponent $H_\kappa$ is given in the figure caption. We note that radial trajectories usually have increasing trends following the trend at which the hyperbolic disc radius increases (Fig.~\ref{hyperbolic_discs}). Trends may also exist in the expected degree and similarity trajectories, although they are often less apparent.

\begin{figure*}
\centering
\includegraphics[width=17.5cm]{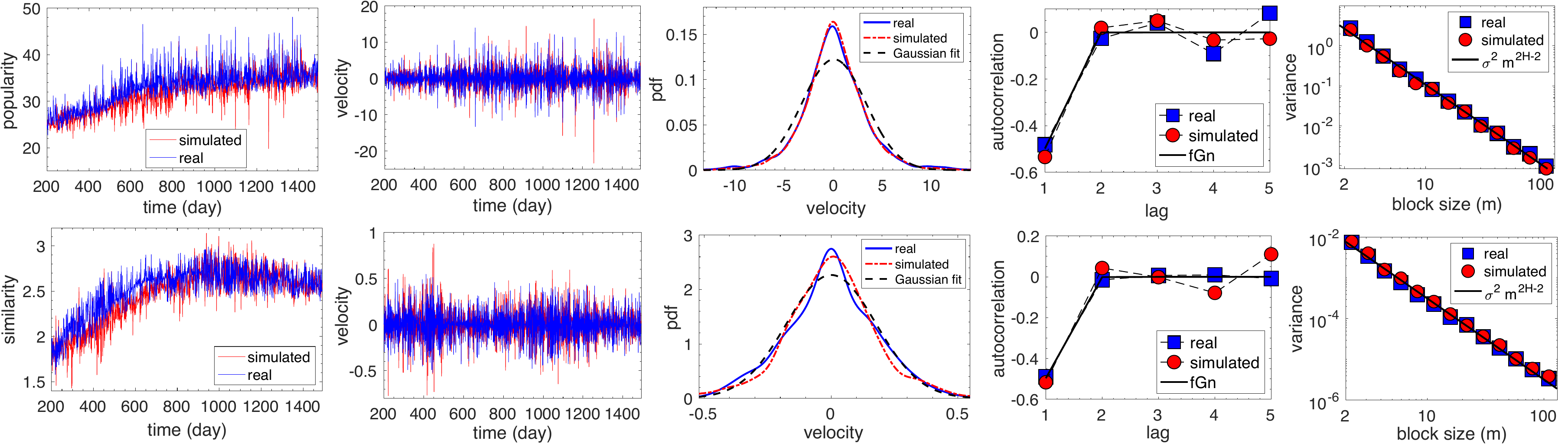}
\vspace{-0.3cm}
\caption{BTC, node 146036. $H_\textnormal{pop}=0.002$, $H_\textnormal{sim}=0.005$.}
~\\
\includegraphics[width=17.5cm]{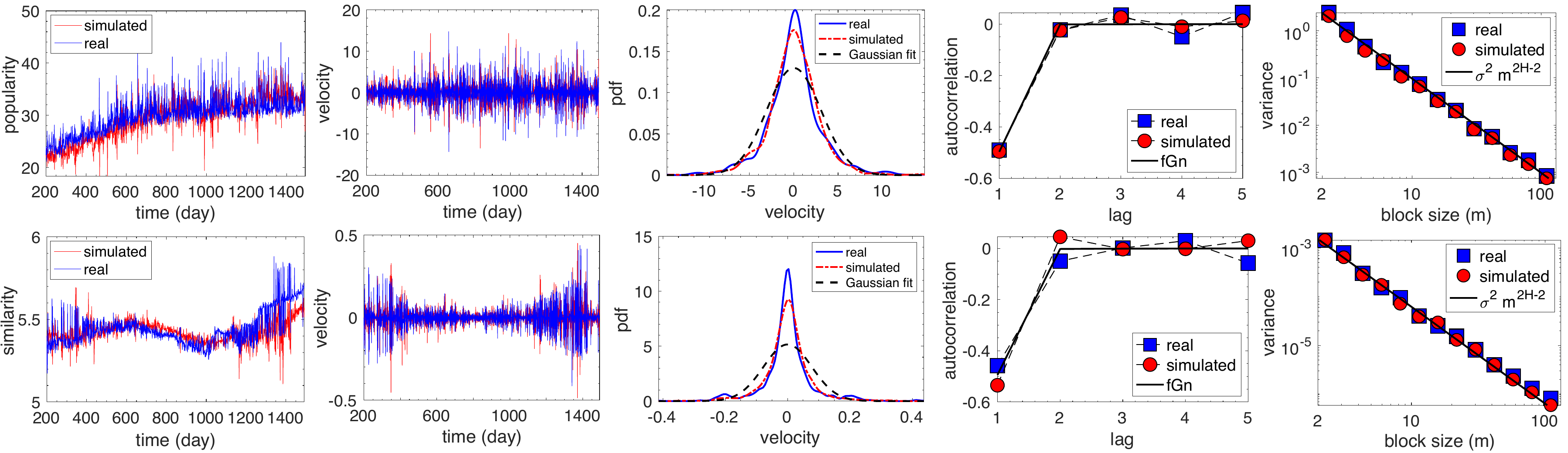}
\vspace{-0.3cm}
\caption{BTC, node 113864. $H_\textnormal{pop}=0.004$, $H_\textnormal{sim}=0.01$.}
~\\
\includegraphics[width=17.5cm]{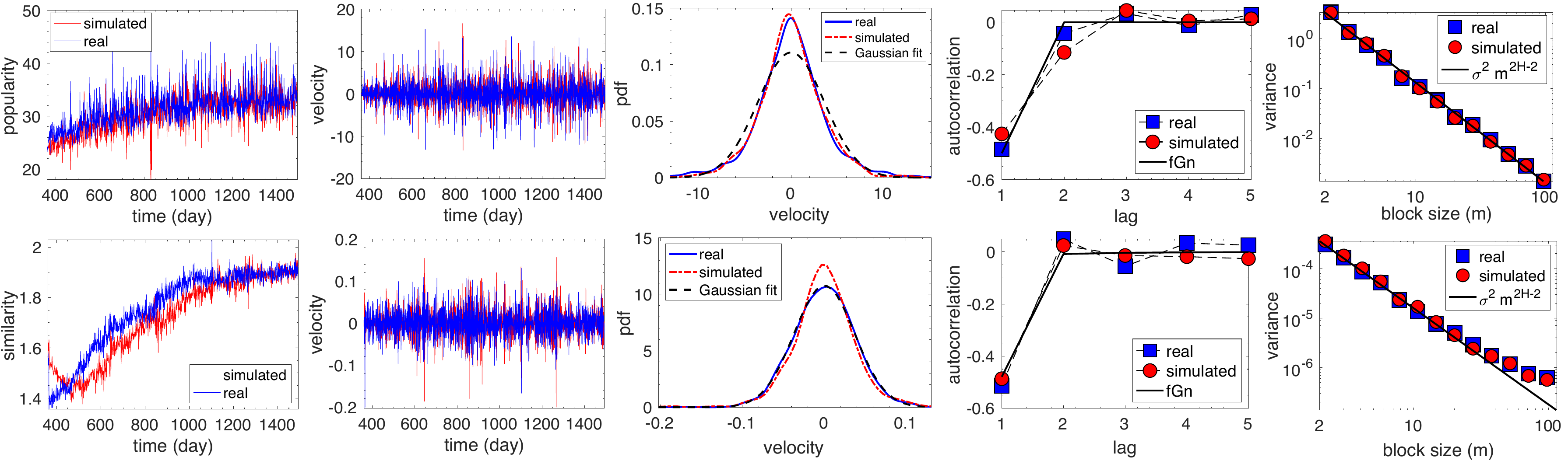}
\vspace{-0.3cm}
\caption{BTC, node 1097470. $H_\textnormal{pop}=0.0001$, $H_\textnormal{sim}=0.03$.}
~\\
\includegraphics[width=17.5cm]{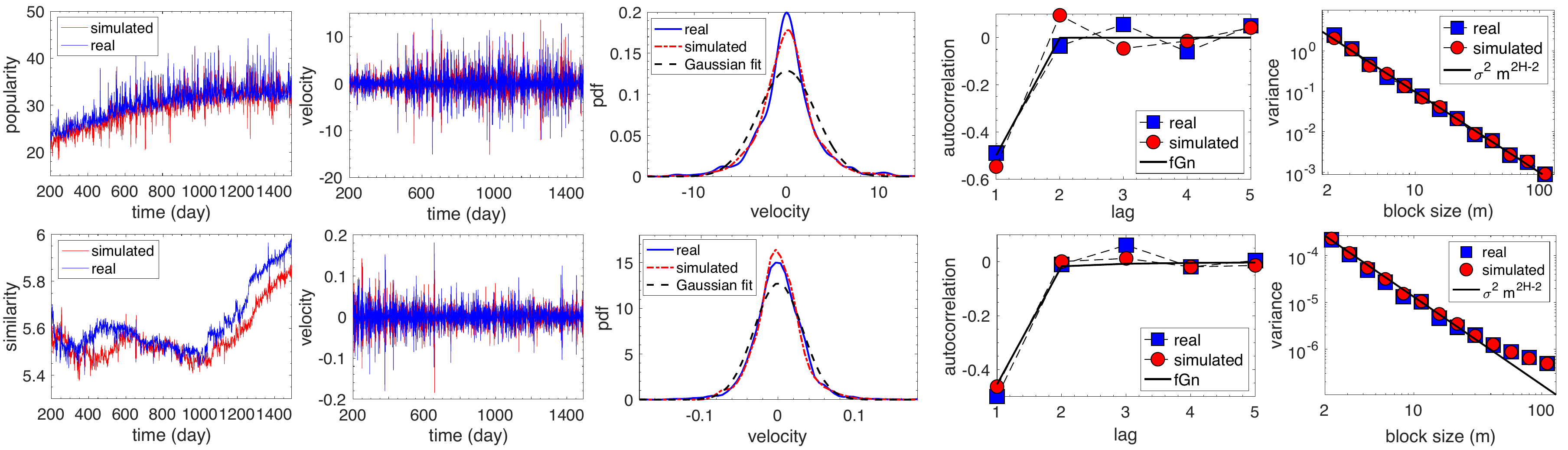}
\vspace{-0.3cm}
\caption{BTC, node 30238. $H_\textnormal{pop}=0.002$, $H_\textnormal{sim}=0.06$.}
\end{figure*}

\begin{figure*}
\centering
\includegraphics[width=17.5cm]{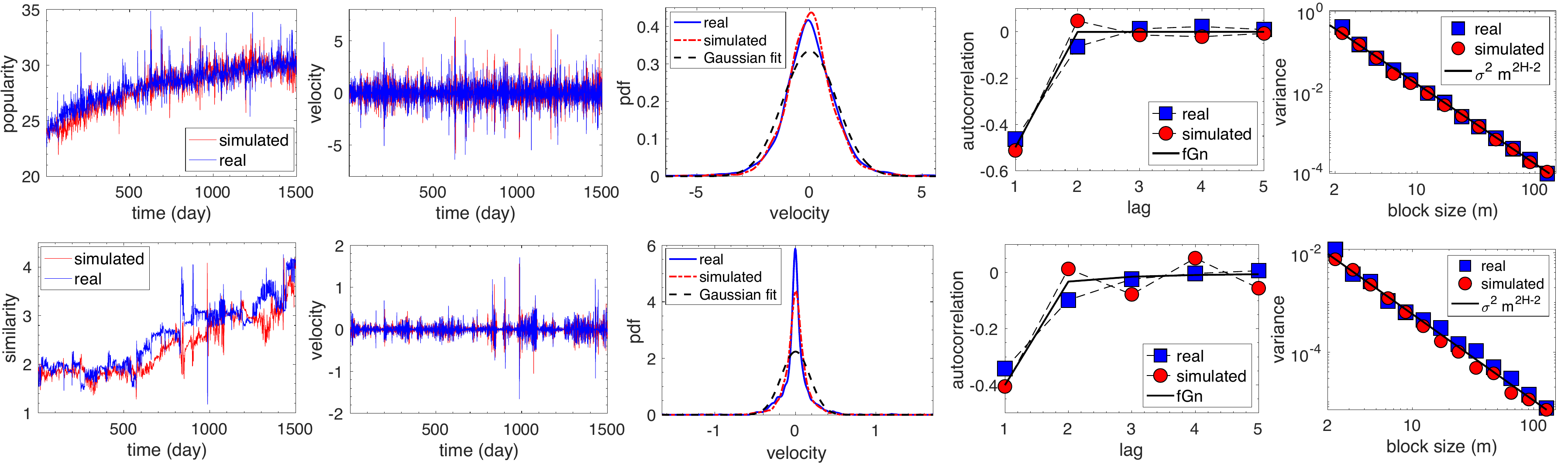}
\vspace{-0.3cm}
\caption{PGP, node 0x0D62001B. $H_\textnormal{pop}=0.01$, $H_\textnormal{sim}=0.13$.}
~\\
\includegraphics[width=17.5cm]{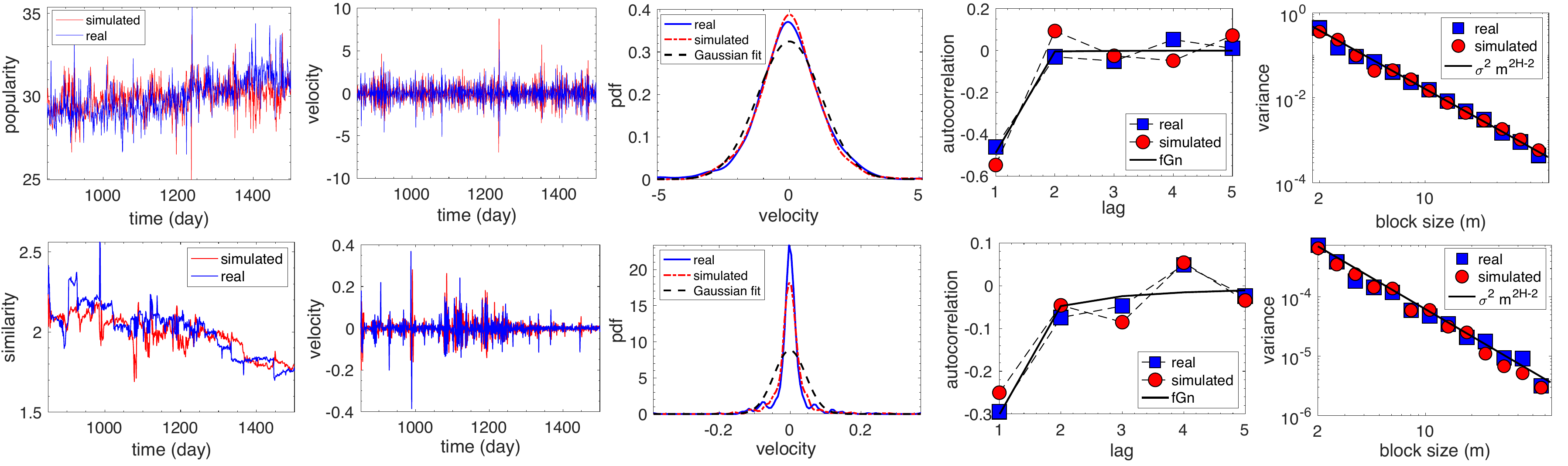}
\vspace{-0.3cm}
\caption{PGP, node 0x0A84BCF5. $H_\textnormal{pop}=0.02$, $H_\textnormal{sim}=0.24$.}
~\\
\includegraphics[width=17.5cm]{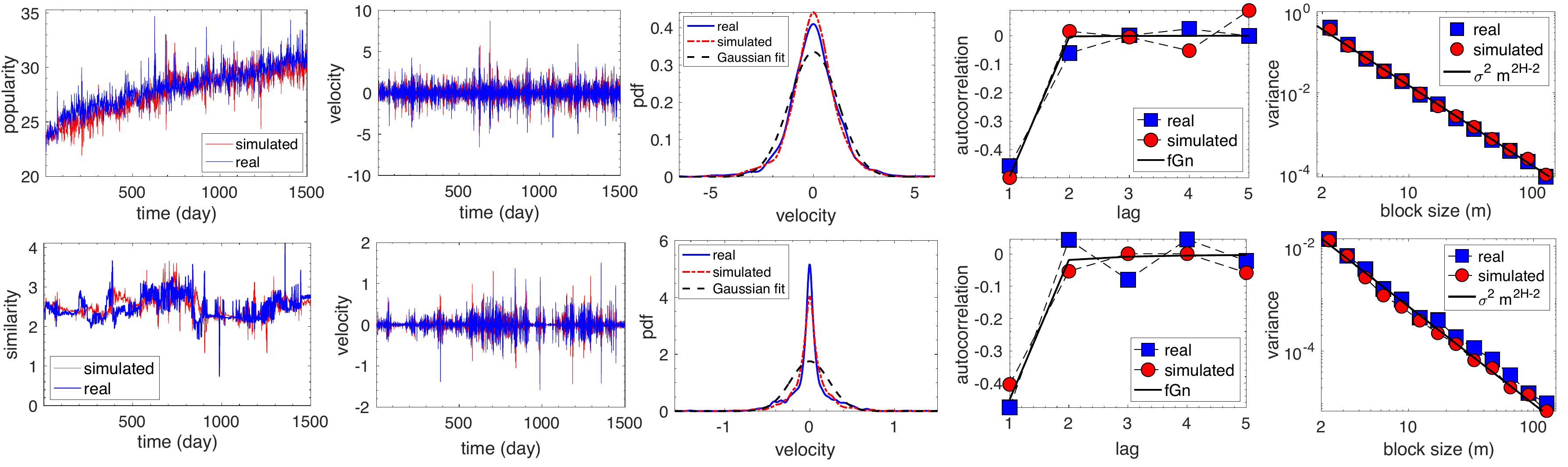}
\vspace{-0.3cm}
\caption{PGP, node 0x0CD8EF97. $H_\textnormal{pop}=0.01$, $H_\textnormal{sim}=0.07$.}
~\\
\includegraphics[width=17.5cm]{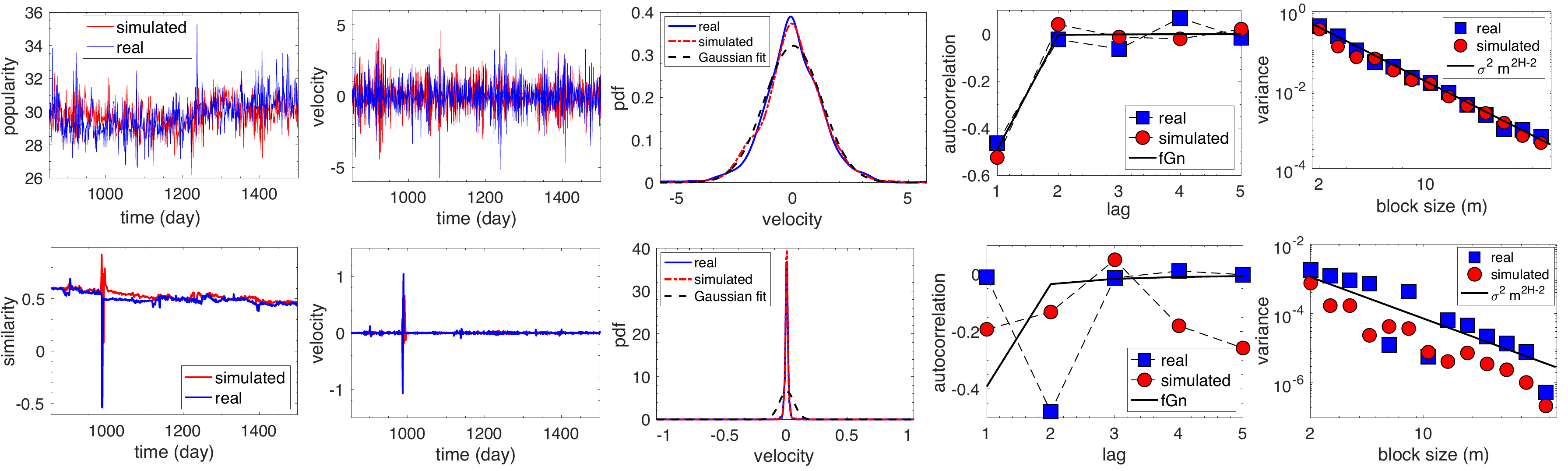}
\vspace{-0.3cm}
\caption{PGP, node 0x0A23A932. $H_\textnormal{pop}=0.01$, $H_\textnormal{sim}=0.14$.}
\end{figure*}

\begin{figure*}
\centering
\includegraphics[width=17.5cm]{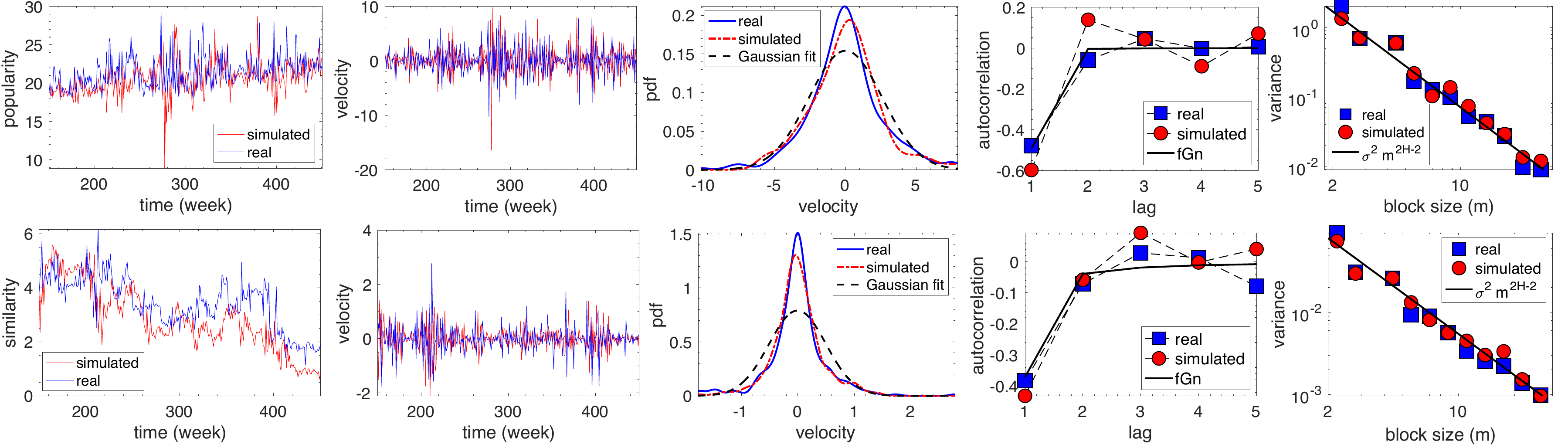}
\vspace{-0.3cm}
\caption{IPv6, node 1239 (Sprint). $H_\textnormal{pop}=0.01$, $H_\textnormal{sim}=0.17$.}
~\\
\includegraphics[width=17.5cm]{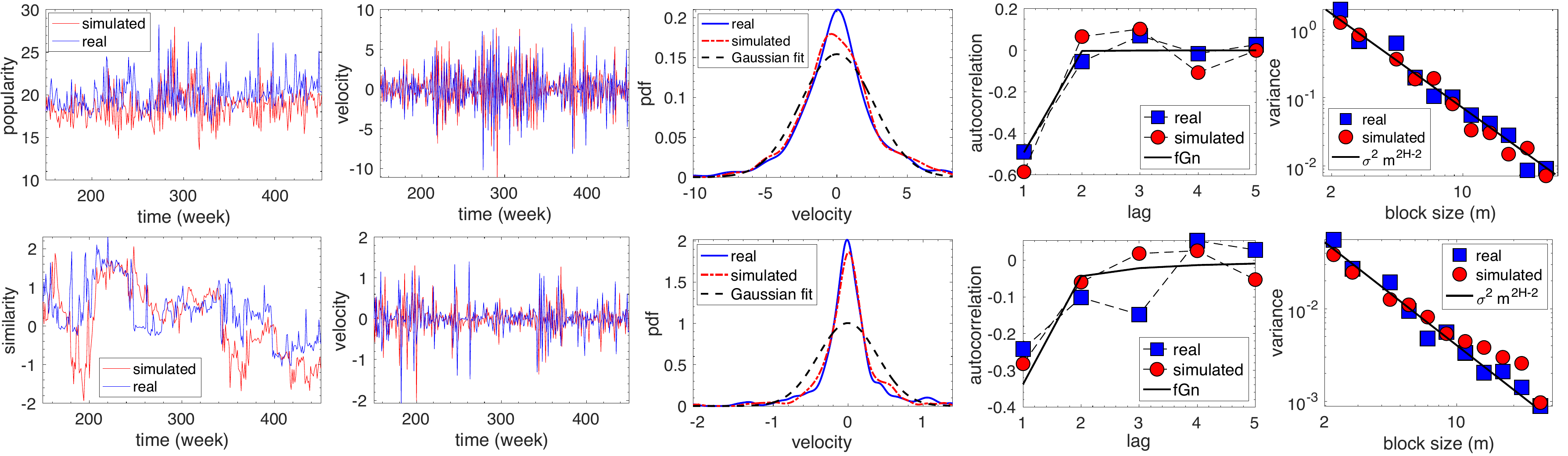}
\vspace{-0.3cm}
\caption{IPv6, node 3320 (Deutsche Telekom). $H_\textnormal{pop}=0.01$, $H_\textnormal{sim}=0.20$.}
~\\
\includegraphics[width=17.5cm]{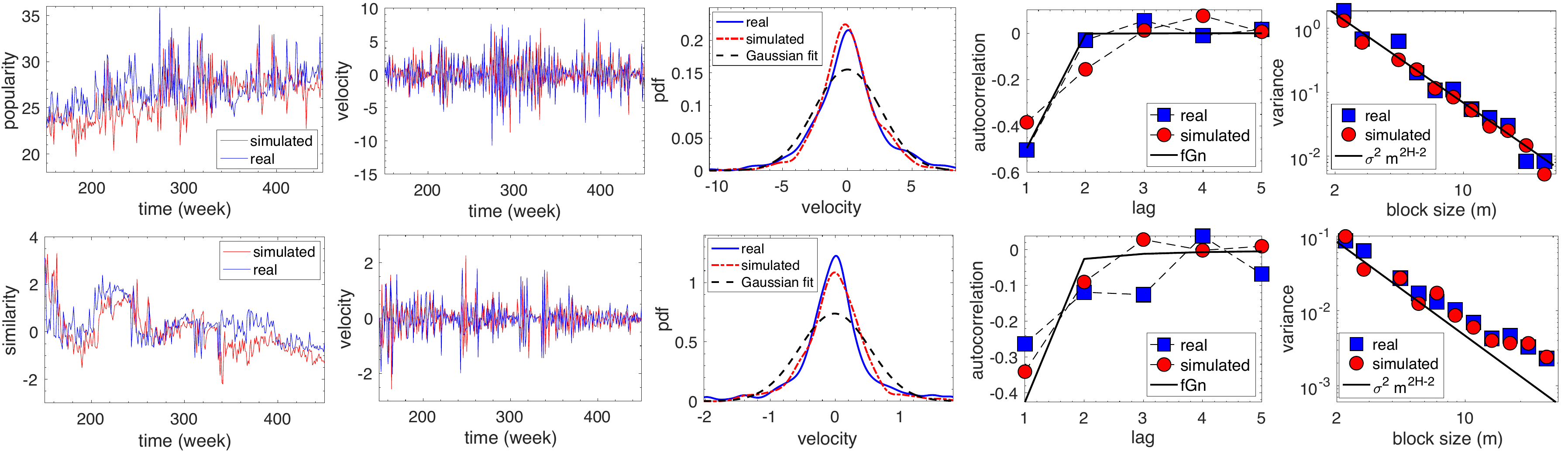}
\vspace{-0.3cm}
\caption{IPv6, node 5539 (SpaceNet). $H_\textnormal{pop}=0.007$, $H_\textnormal{sim}=0.10$.}
~\\
\includegraphics[width=17.5cm]{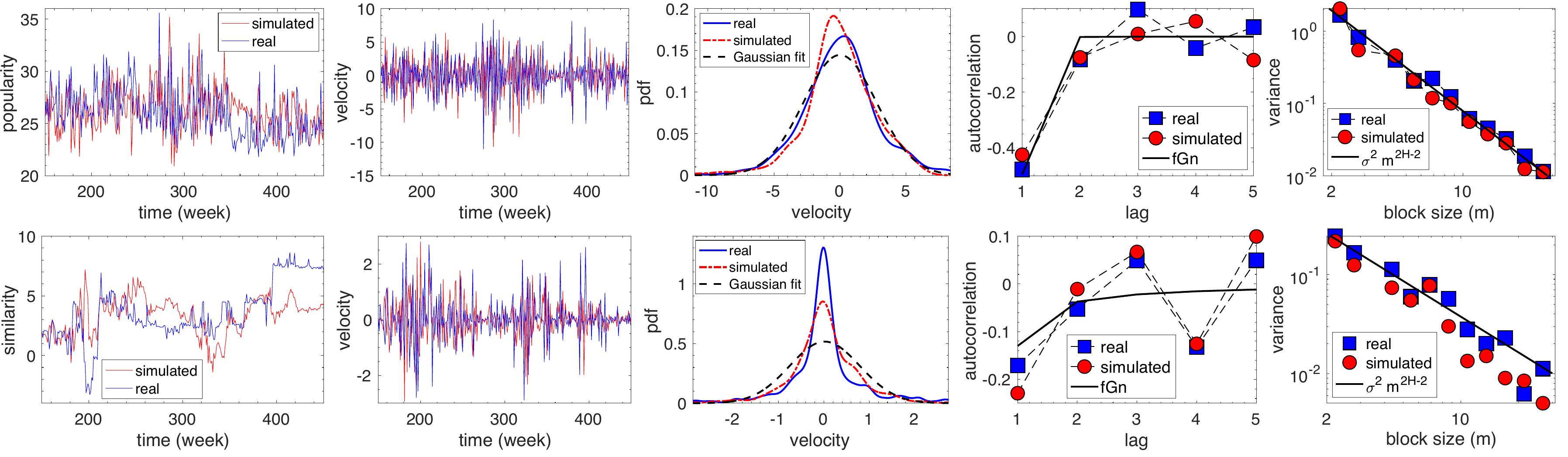}
\vspace{-0.3cm}
\caption{IPv6, node 14840 (BR Digital). $H_\textnormal{pop}=0.006$, $H_\textnormal{sim}=0.40$.}
\end{figure*}

\begin{figure*}
\centering
\includegraphics[width=17.5cm]{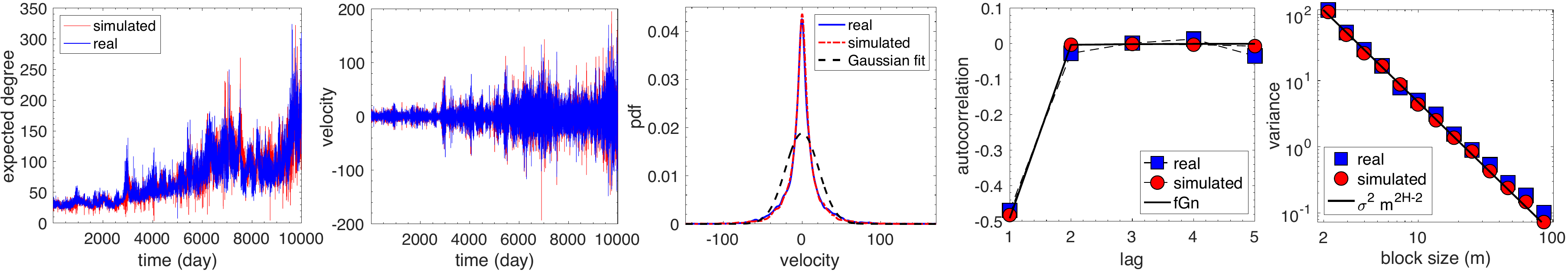}
\vspace{-0.3cm}
\caption{US Air, node SAN. $H_\kappa=0.01$.}
\label{USAir_kappa}
~\\
\includegraphics[width=17.5cm]{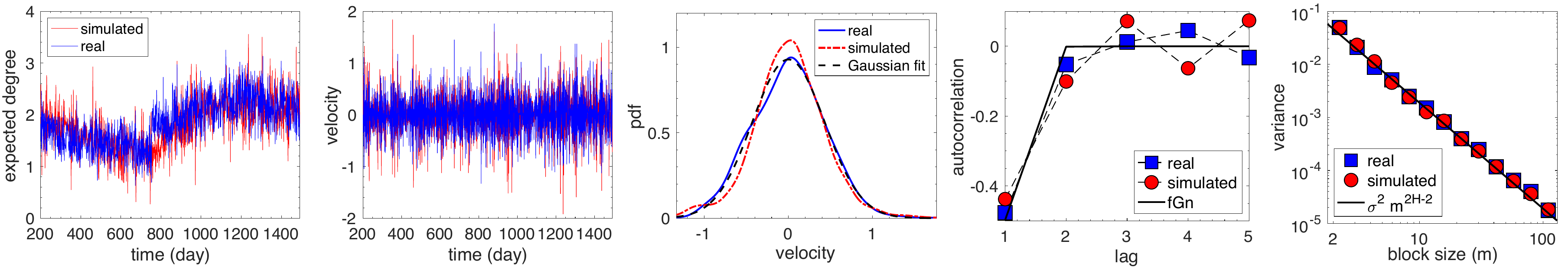}
\vspace{-0.3cm}
\caption{BTC, node 113864. $H_\kappa=0.004$.}
~\\
\includegraphics[width=17.5cm]{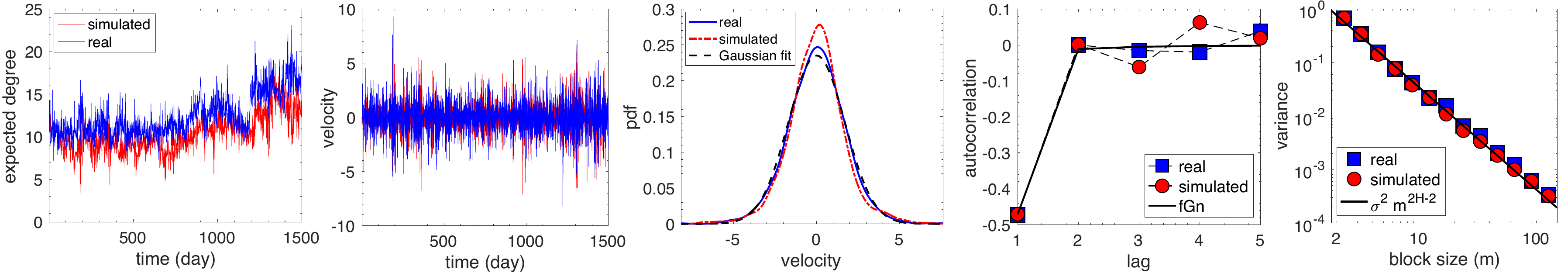}
\vspace{-0.3cm}
\caption{PGP, node 0x0D62001B. $H_\kappa=0.04$.}
~\\
\includegraphics[width=17.5cm]{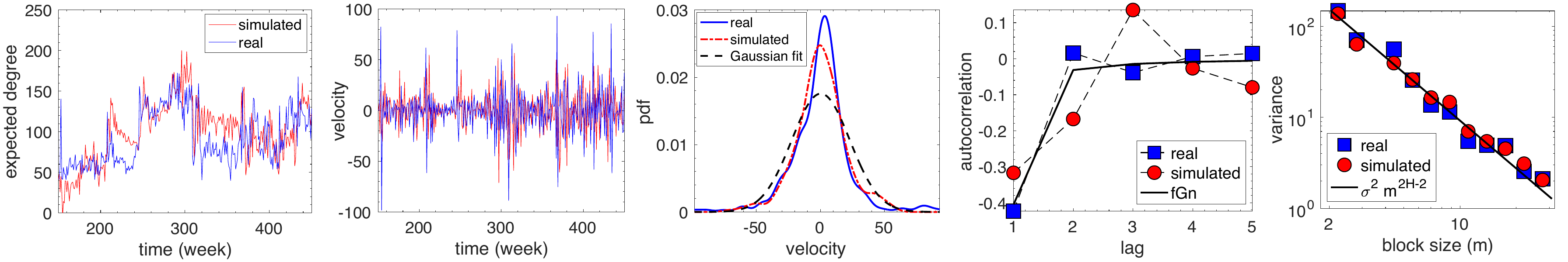}
\vspace{-0.3cm}
\caption{IPv6, node 1239. $H_\kappa=0.13$.}
~\\
\includegraphics[width=17.5cm]{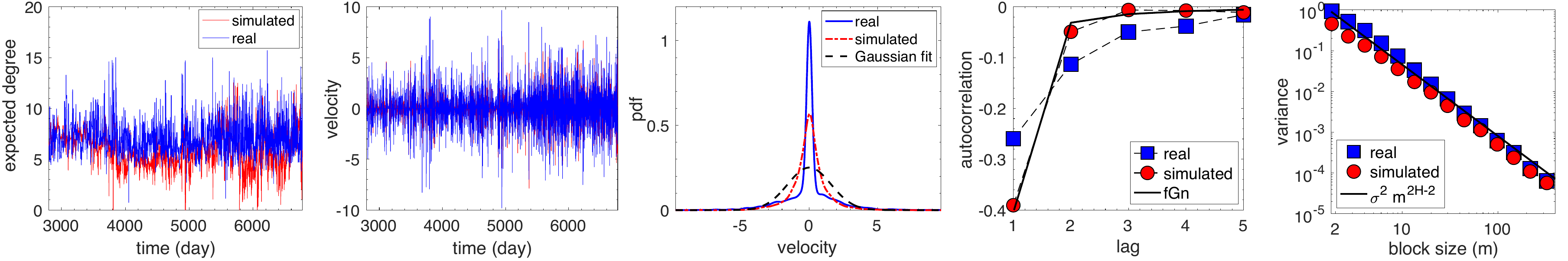}
\vspace{-0.3cm}
\caption{arXiv, node 1058. $H_\kappa=0.13$.}
\label{arXiv_kappa}
\end{figure*}

\section{A glimpse on predictability}
\label{sec:pred_examples}

Appendix~\ref{pop_sim_examples} shows that we can construct simulated trajectories resembling the real ones, using the model of Appendix~\ref{sec:fbm_model}, i.e., Eq.~(\ref{eq:oursim}). In principle, the same model can be used for predictions, i.e., for predicting the future evolution of the trajectories, such as the area in which the trajectories are expected to be located and their expected trend. To this end, one will need to make an educated guess for the model's parameters during the prediction period, i.e., for the values of $H$, $\mu_i$, and $\sigma_i$, based on historical data. 

\begin{figure*}
\centering
\includegraphics[width=17.5cm]{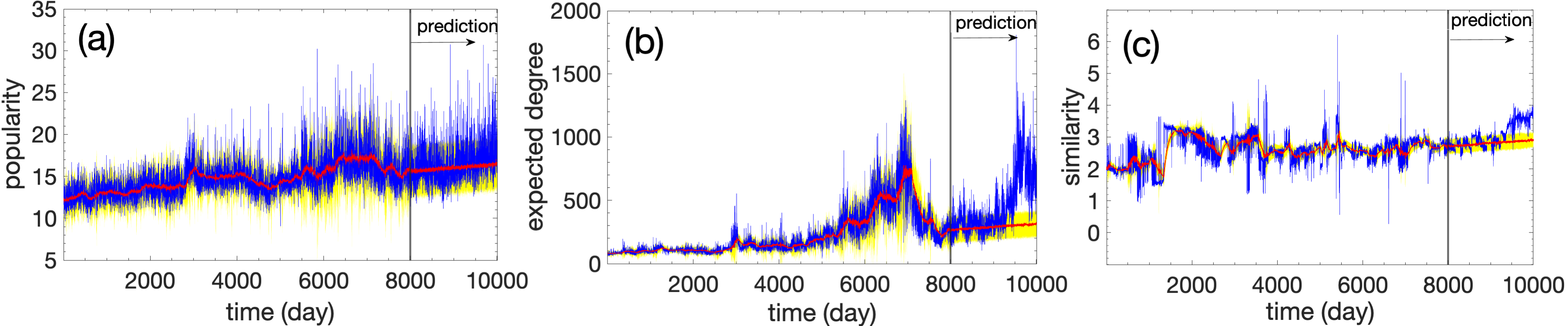}
\vspace{-0.3cm}
\caption{Prediction of the popularity-similarity trajectories of node LAX in the US Air. The real trajectories are shown in blue. The red line is the average across $100$ simulated trajectories constructed using the model of Appendix~\ref{sec:fbm_model}. In (a) and (b) the yellow-shaded area stands for two standard deviations away from the average. In (c) it stands for one standard deviation. The model parameters for the prediction period are estimated using the data up to the vertical line (see text). Predictions start immediately after the vertical line. 
\label{USAir_pred}}
~\\
\includegraphics[width=17.5cm]{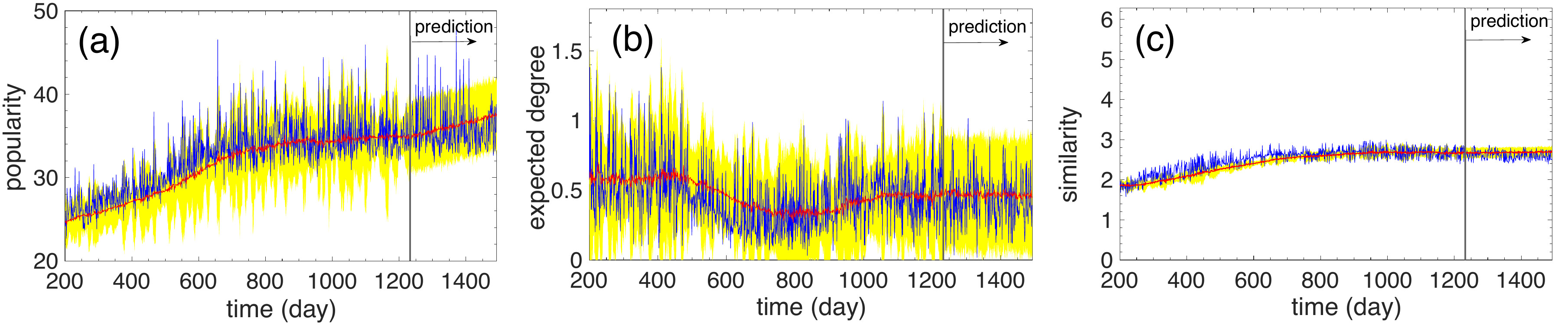}
\vspace{-0.3cm}
\caption{Same as in Fig.~\ref{USAir_pred}, but for node 146036 in BTC.
\label{BTC_pred}}
~\\
\includegraphics[width=17.5cm]{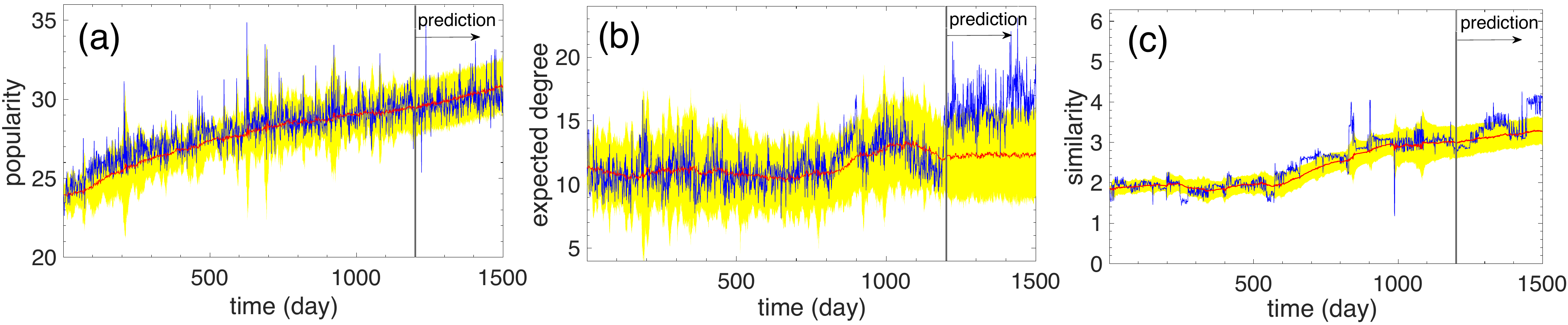}
\vspace{-0.3cm}
\caption{Same as in Fig.~\ref{USAir_pred}, but for node 0x0D62001B in PGP WoT.
\label{PGP_pred}}
~\\
\includegraphics[width=17.5cm]{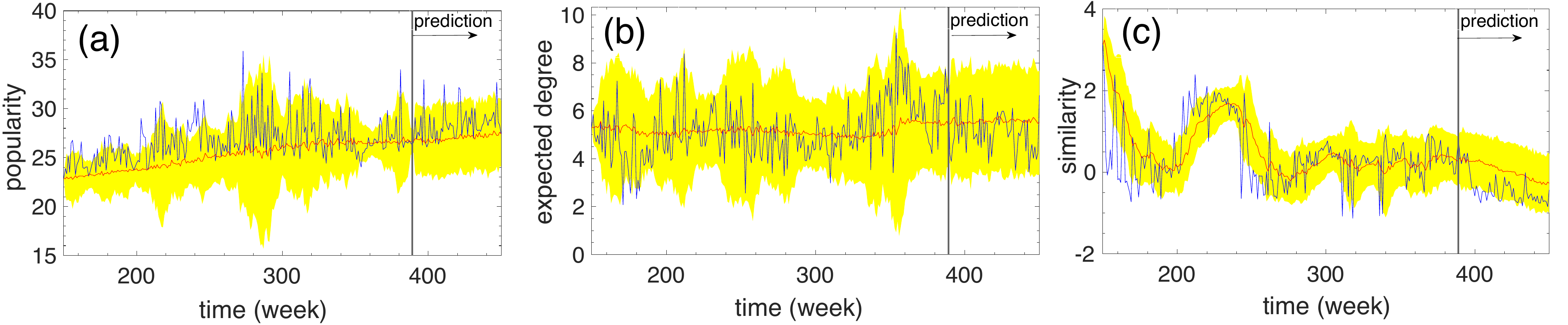}
\vspace{-0.3cm}
\caption{Same as in Fig.~\ref{USAir_pred}, but for node 5539 in IPv6.
\label{IPv6_pred}}
~\\
\includegraphics[width=17.5cm]{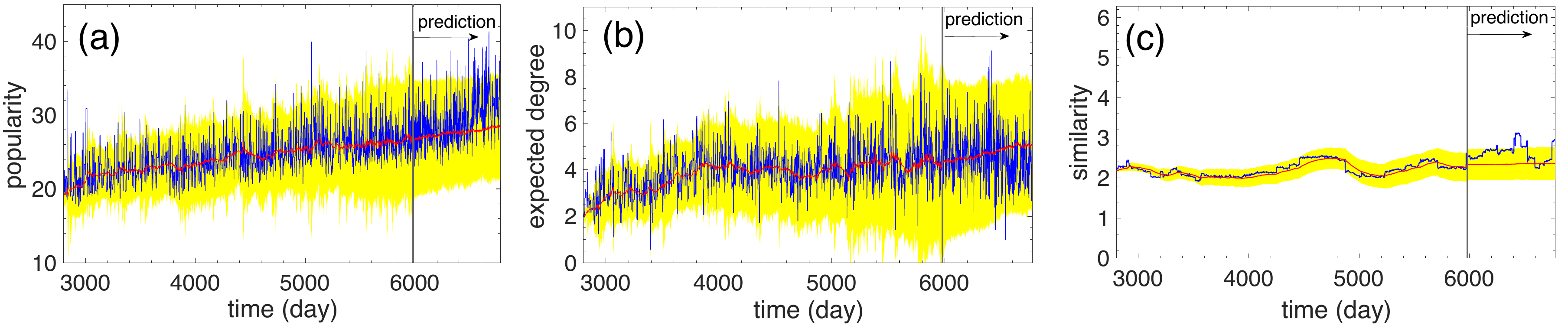}
\vspace{-0.3cm}
\caption{Same as in Fig.~\ref{USAir_pred}, but for node 1674 in arXiv.
\label{arXiv_pred}}
\end{figure*}

Figures~\ref{USAir_pred}-\ref{arXiv_pred} illustrate this idea. Here we use the first $80$\% of each trajectory as historical data, where we estimate $H$, $\mu_i$, and $\sigma_i$, as described in Appendix~\ref{sec:fbm_model}. Then, we predict the evolution of the subsequent $20$\% of each trajectory. For the predictions we simply use the historical estimate of $H$, a constant volatility $\sigma_i=\sigma$ equal to the average of the historical $\sigma_i$, and a constant trend $\mu_i=\mu$ equal to the average of the historical $\mu_i$. (In Fig.~\ref{BTC_pred}(c), $\mu$ is equal to the average of $\mu_i$ from the preceding year, rather than encompassing the entire historical range of $\mu_i$.) We find that in several cases this simple approach yields reasonable prediction results, as in Figs.~\ref{USAir_pred}-\ref{arXiv_pred}. Identifying the best automated approaches for fine-tuning the model's parameters for predictions and performing a comprehensive evaluation of the  model's predictive capabilities are beyond the scope of this paper.

\section{Time-scale considerations}
\label{sec:timescales}

Real world networks are continuous-time dynamical systems. Therefore, in our analysis we have considered daily snapshots of the real-world networks, which is the finest level of time resolution that we can have from the available data repositories. The only exception is in the case of the IPv6 Internet, where we have considered weekly snapshots; we did this in order to avoid potential link fluctuations that are due to CAIDA's measurement methodology~\cite{as_topo_data_ipv6}.

\begin{figure*}
\includegraphics[width=17.5cm]{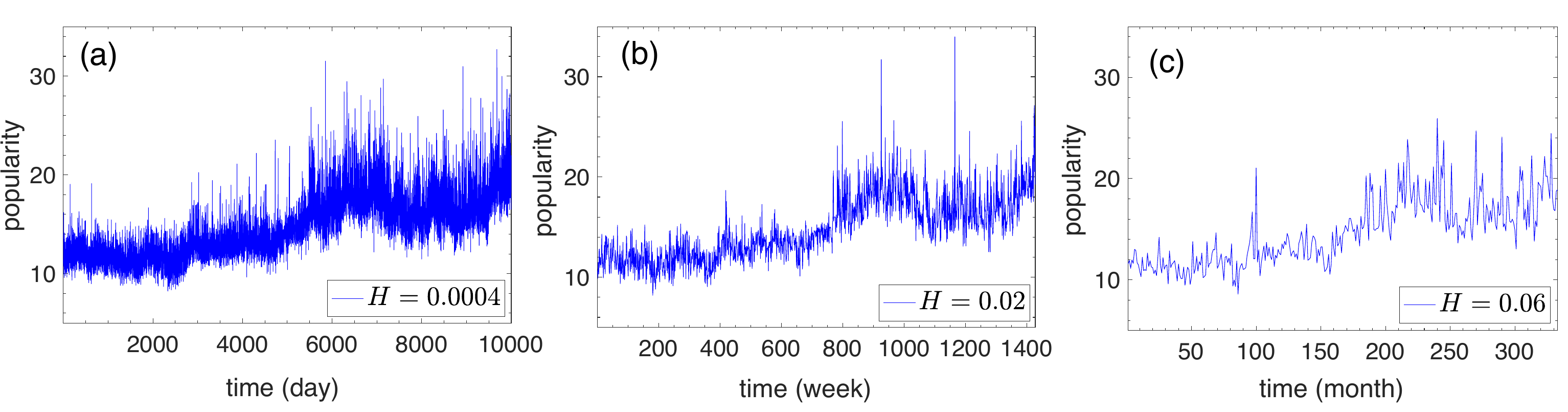}
\caption{Radial popularity trajectory of Charlotte Douglas International Airport (CLT) in the US Air transportation network. (a) For the case of daily snapshots, (b) for the case of weekly snapshots, and (c) for the case of monthly snapshots. The estimated Hurst exponent $H$ in each case is indicated in the legend. The average Hurst exponent across all radial trajectories for the case of weekly and monthly snapshots is respectively $0.03$ and $0.08$ (vs. $0.004$ for daily snapshots).
\label{fig:CLT_pop_aggr}}
\vspace{-3cm}
\end{figure*}

\begin{figure*}
\includegraphics[width=17.5cm]{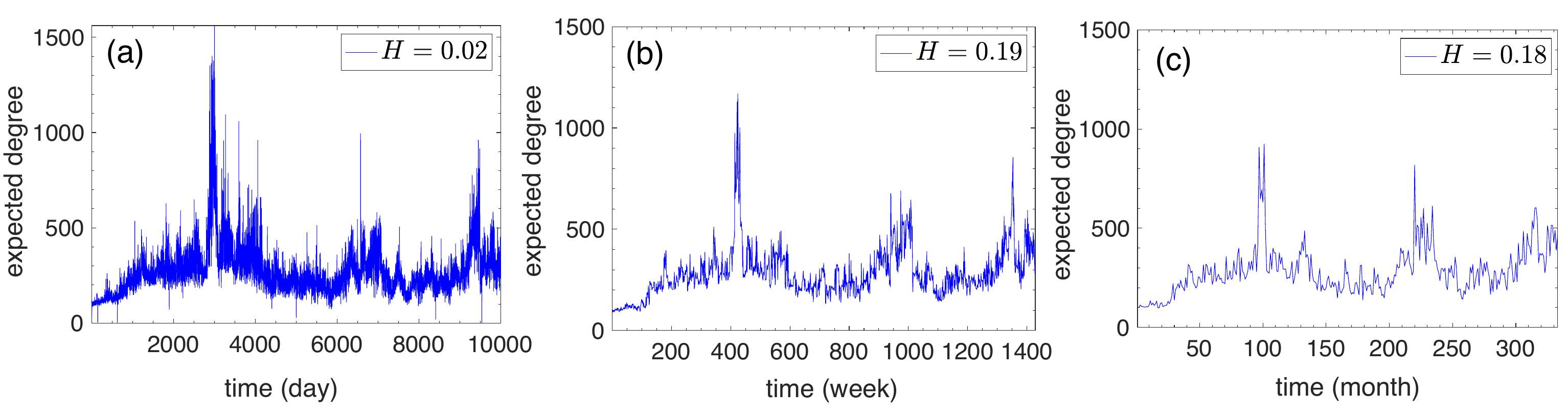}
\caption{Same as in Fig.~\ref{fig:CLT_pop_aggr}, but for CLT's expected degree trajectory. The average Hurst exponent across all degree trajectories  for weekly and monthly snapshots is respectively $0.14$ and $0.20$ (vs. $0.01$ for daily snapshots).
\label{fig:CLT_deg_aggr}}
\vspace{-3cm}
\end{figure*}

\begin{figure*}
\includegraphics[width=17.5cm]{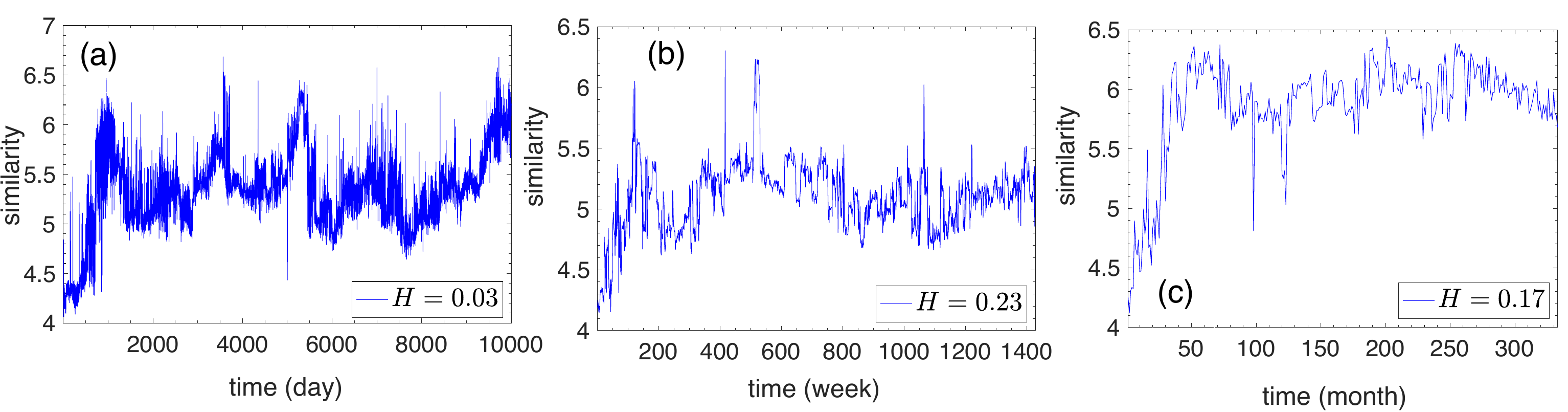}
\caption{Same as in Fig.~\ref{fig:CLT_pop_aggr}, but for CLT's similarity trajectory. The average Hurst exponent across all similarity trajectories  for weekly and monthly snapshots is respectively $0.23$ and $0.18$ (vs. $0.08$ for daily snapshots).
\label{fig:CLT_sim_aggr}}
\end{figure*}

Aggregating network snapshots over longer time intervals, such as weekly or monthly intervals, and studying the nodes' trajectories in the hyperbolic embeddings of the time-aggregated snapshots, yields qualitatively similar results, cf. Figs.~\ref{fig:CLT_pop_aggr}-\ref{fig:CLT_sim_aggr}. However, as it is evident by Figs.~\ref{fig:CLT_pop_aggr}-\ref{fig:CLT_sim_aggr}, coarser aggregation misses finer-grained dynamics and can influence the Hurst exponent of the trajectories.

\section{Popularity-similarity trajectories in geometric vs. non-geometric networks}
\label{sec:geom_vs_nongeom}

In this section we provide a detailed description of the temporal network model considered in the main text, along with supplementary figures. The model is based on the $\mathbb{S}^{1}$ model of complex networks~\cite{Krioukov2010}, which we overview first.

\subsection{$\mathbb{S}^{1}$ model}
\label{sec:s1_model}

Each node $i$ in the $\mathbb{S}^{1}$ model has hidden variables $\kappa_i, \theta_i$. The hidden variable $\kappa_i$ is the node's expected degree in the resulting network, while $\theta_i$ is the angular (similarity) coordinate of the node on a circle of radius $\hat{R}=N/(2\pi)$, where $N$ is the total number of nodes. To construct a network that has size $N$, average node degree $\bar{k}$, power law degree distribution with exponent $\gamma > 2$, and temperature $T \in (0,1)$, we perform the following steps:
\begin{enumerate}
\item[i.] Sample the angular coordinates of nodes $\theta_i$, $i=1,2,\ldots,N$, uniformly at random from $[0, 2\pi]$, and their hidden variables $\kappa_{i}$, $i=1,2,\ldots,N$, from the probability density function
\begin{equation}
\rho(\kappa) = (\gamma-1) \kappa_0^{\gamma-1} \kappa^{-\gamma},
\end{equation}
where $\kappa_0=\bar{k}(\gamma-2)/(\gamma-1)$ is the expected minimum node degree;
\item[ii.] Connect every pair of nodes $i,j$ with probability
\begin{equation}
\label{eq:p_s1}
p(\chi_{ij})=\frac{1}{1+\chi_{ij}^{1/T}},
\end{equation}
where $\chi_{ij}= \hat{R} \Delta \theta_{ij}/(\mu\kappa_i\kappa_j)$ is the effective distance between $i$ and $j$, $\Delta \theta_{ij}$ is the angular distance, and $\mu =\sin{T \pi}/(2\bar{k}T\pi)$ ensures that the expected degree in the network is $\bar{k}$.
\end{enumerate}
Smaller values of  the temperature $T$ favor connections at smaller effective distances and increase the average clustering in the network, which decreases to zero with $T \in (0, 1)$. Other forms of $\rho(\kappa)$ can also be used in the model. The model is isomorphic to random hyperbolic graphs (RHGs) after transforming the expected degrees $\kappa_i$ to radial coordinates $r_i$ via $r_i = R - 2 \ln{(\kappa_i/\kappa_0)}$, where $R \sim 2 \ln{N}$ is the radius of the hyperbolic disc~\cite{Krioukov2010}. 

It can be shown~\cite{Krioukov2010} that the configuration model, i.e., the ensemble of graphs with given expected degrees~\cite{ChungLu2002, Newman2004}, is an infinite temperature limit of the $\mathbb{S}^{1}$ model, where the connection probability in~(\ref{eq:p_s1}) becomes
\begin{equation}
\label{eq:p_cm}
p(\kappa_i, \kappa_j) = \frac{1}{1+N \bar{k}/(\kappa_i \kappa_j)}.
\end{equation}
In this limit, only the nodes' expected degrees matter, while the similarity distances among the nodes are completely ignored. Thus, the model is not geometric. If we further let $\kappa_i = \bar{k}$ for all nodes $i$, then the connection probability in~(\ref{eq:p_cm}) reduces to
\begin{equation}
\label{eq:p_rg}
p=\frac{1}{1+N/\bar{k}}.
\end{equation}
In this case, the nodes' expected degrees do not matter either and the model degenerates to classical random graphs~\cite{SoRa51}, where each pair of nodes is connected with the same probability and the resulting degree distribution is Poissonian. Both in the configuration model and in random graphs clustering is asymptotically (as $N \to \infty$) zero. 

\subsection{Link rewirings and popularity-similarity trajectories}
\label{sec:temp_model}

In the main text we consider the following simple model of network snapshots $G_t, t=1, 2, \dots, \tau$, that undergo link rewirings. In the geometric version, snapshot $G_1$ is constructed according to the $\mathbb{S}^{1}$ model (RHGs). Then, each subsequent snapshot $G_t$ is obtained from $G_{t-1}$ as follows. First, we randomly delete a number of links $l$ from $G_{t-1}$. Then, we add $l$ links among node pairs according to the probability in~(\ref{eq:p_s1}), while ensuring that no multi-edges are created and that the average clustering is preserved. The described link rewiring effectively creates snapshots $G_t$ with the same statistical properties, including the same degree distribution. Thus, it mimics behavior in real networks, where statistical quantities, such as average degree, shape of degree distribution, clustering strength, etc., are approximately preserved over time or change slowly (cf. Figs.~\ref{USAir_props}-\ref{arXiv_props}(b),(c)).    

In the non-geometric version of the model, snapshot $G_1$ is constructed either according to the configuration model (CM) or to random graphs (RGs), and the addition of links in subsequent snapshots is done according to~(\ref{eq:p_cm}) and~(\ref{eq:p_rg}), respectively. We use the following parameters: $N=300$, $\bar{k}=8$, $\tau=1000$, $l=30$, $\gamma=2.5$, and $T=0.1$. We note that the percentage of link rewirings in each step ($2.5$\%) is of the same order as in some of the considered real networks---the US Air and IPv6 (Appendix~\ref{sec:data}). We also note that for the geometric networks the average clustering is $0.79$, which is strong as in real networks (cf. Figs.~\ref{USAir_props}-\ref{arXiv_props}(c)). As mentioned in the main text, once the temporal networks are created, we embed them into hyperbolic spaces and obtain the nodes' popularity and similarity trajectories following the same procedure as in the real networks. 
 
\begin{figure*}
\centering
\includegraphics[width=16.5cm]{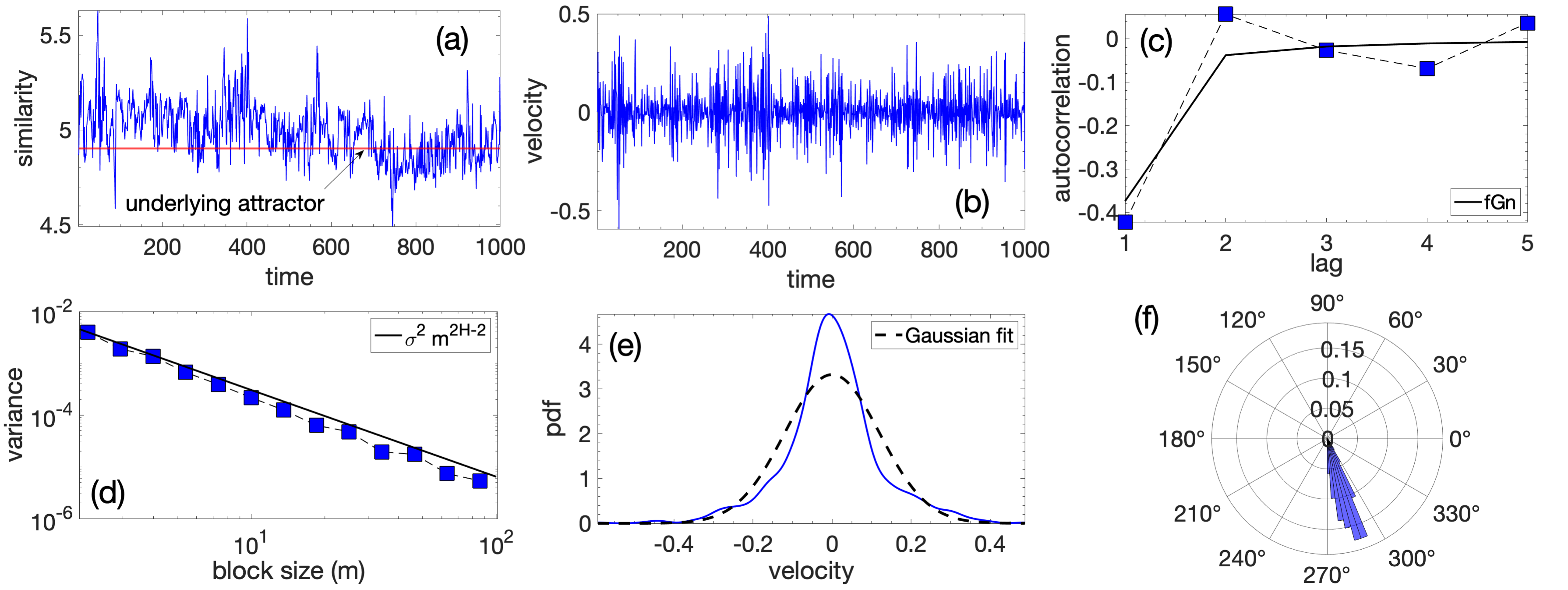}
\caption{Similarity trajectory in a temporal network constructed according to the $\mathbb{S}^{1}$ model (RHGs). Plots (a)-(f) show the same as in Fig.~\ref{sim_clt}(a)-(f) in the main text. The red line in~(a) indicates  the node's hidden similarity coordinate $\theta=4.9$. The estimated Hurst exponent for the trajectory is $0.16$.}
\label{sim_s1}
\end{figure*}

Figure~\ref{sim_s1} shows that similarity trajectories in the RHGs model possess similar characteristics as in real networks (the figure corresponds to the trajectory shown in Fig.~\ref{geom_effect}(c) in the main text). These results hold irrespective of the distribution of expected degrees $\rho(\kappa)$. Further, we note that strong anti-persistence in the similarity trajectories can appear at both lower and higher values of the temperature parameter $T$ (corresponding respectively to stronger and weaker clustering). 

\begin{figure*}
\centering
\includegraphics[width=16.5cm]{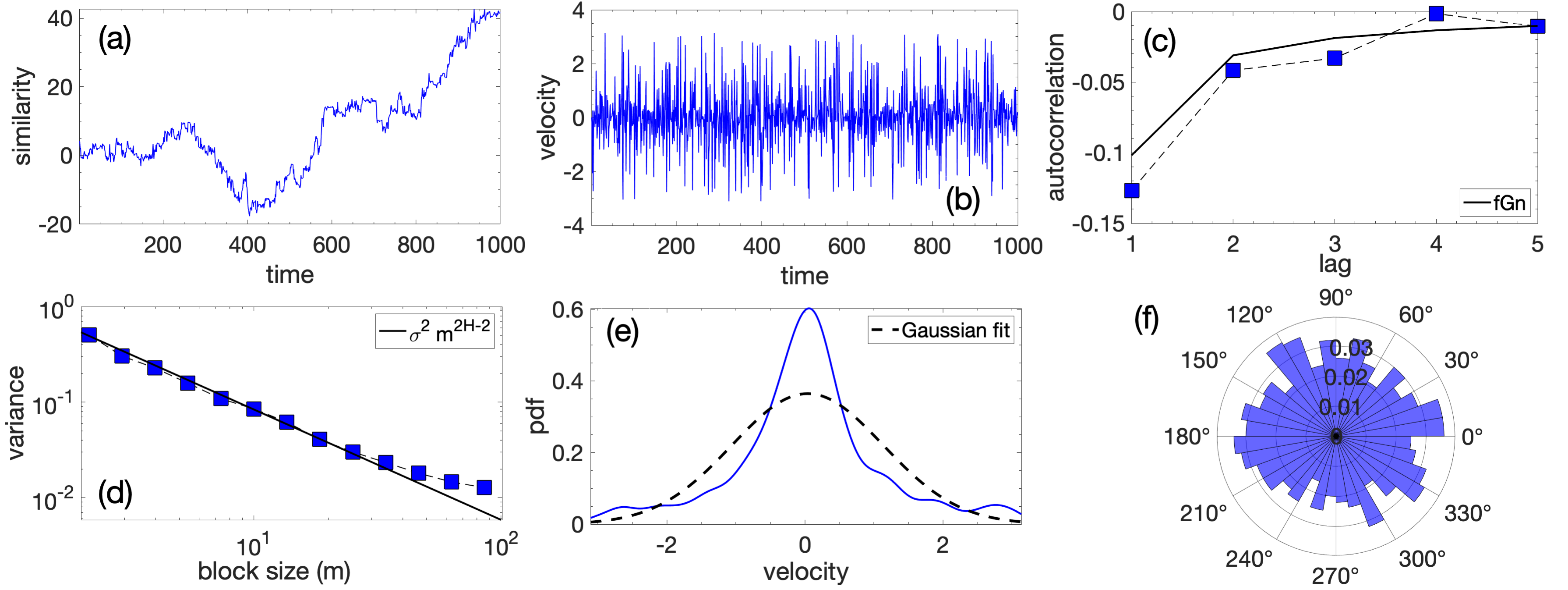}
\caption{Same as in Fig.~\ref{sim_s1}, but for a trajectory in a temporal network constructed according to CM with the same power-law distribution of expected degrees. The estimated Hurst exponent for the trajectory is $0.42$.}
\label{sim_cm}
\end{figure*}

\begin{figure*}
\centering
\includegraphics[width=16.5cm]{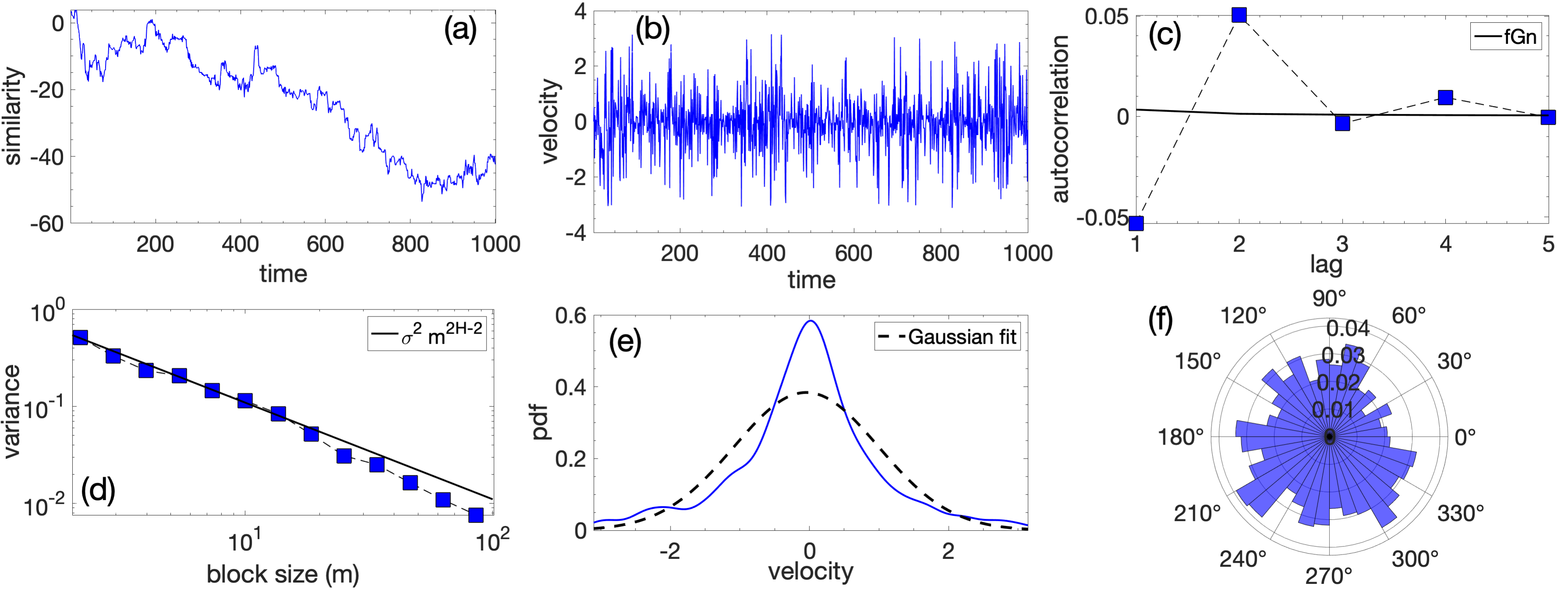}
\caption{Same as in Fig.~\ref{sim_s1}, but for a trajectory in a temporal network constructed according to RGs with the same expected degree. The estimated Hurst exponent for the trajectory is $0.5$.}
\label{sim_rg}
\end{figure*}

Figures~\ref{sim_cm} and~\ref{sim_rg} show corresponding results for the non-geometric version of the model (Fig.~\ref{sim_cm} corresponds to the trajectory shown in Fig.~\ref{geom_effect}(e) in the main text). The distributions of Hurst exponents across all similarity trajectories in the networks of Figs.~\ref{sim_s1}-\ref{sim_rg} are shown in Fig.~\ref{geom_effect}(b) in the main text.  

\begin{figure*}
\centering
\includegraphics[width=16.5cm]{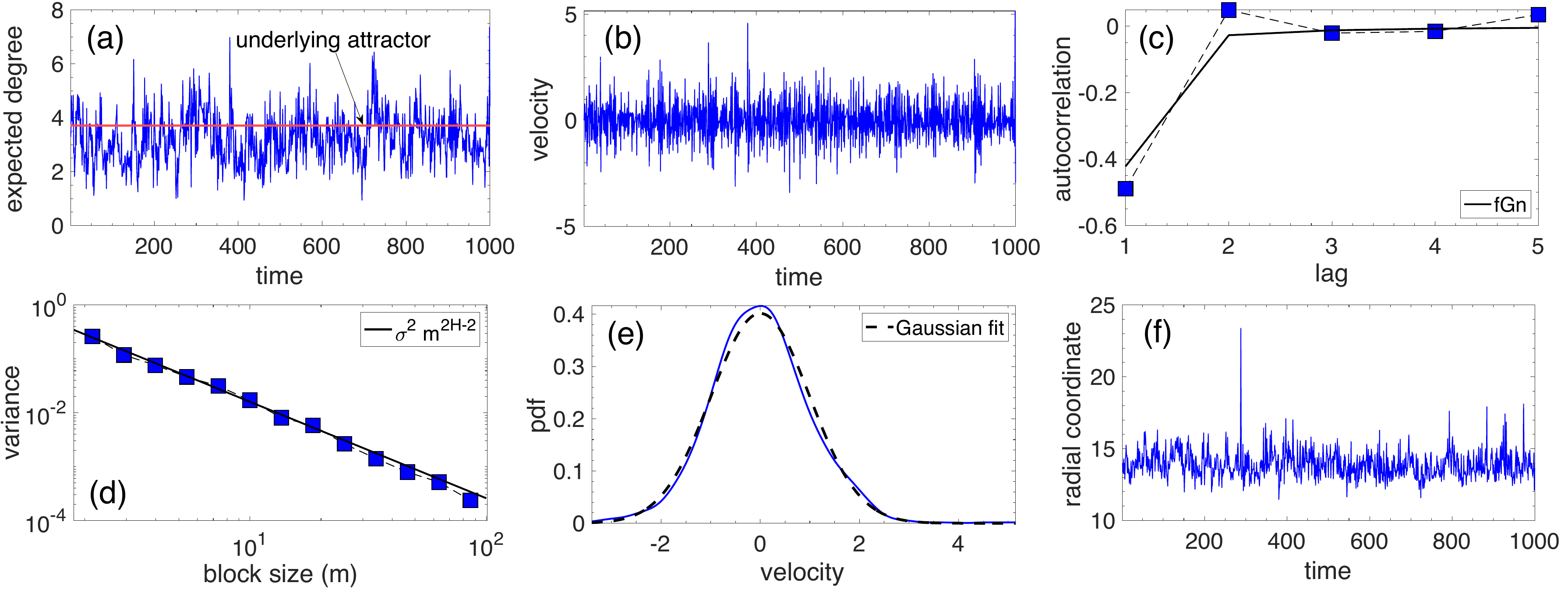}
\caption{Plots (a)-(e) are the same as in Fig.~\ref{sim_s1}(a)-(e), but for the expected degree trajectory of the corresponding node (RHGs model). The red line in~(a) indicates the node's hidden degree variable $\kappa=3.7$. Plot~(f) shows the node's radial trajectory. The estimated Hurst exponents for the degree and radial trajectories are respectively $0.10$ and $0.04$.}
\label{deg_s1}
\end{figure*}

\begin{figure*}
\centering
\includegraphics[width=16.5cm]{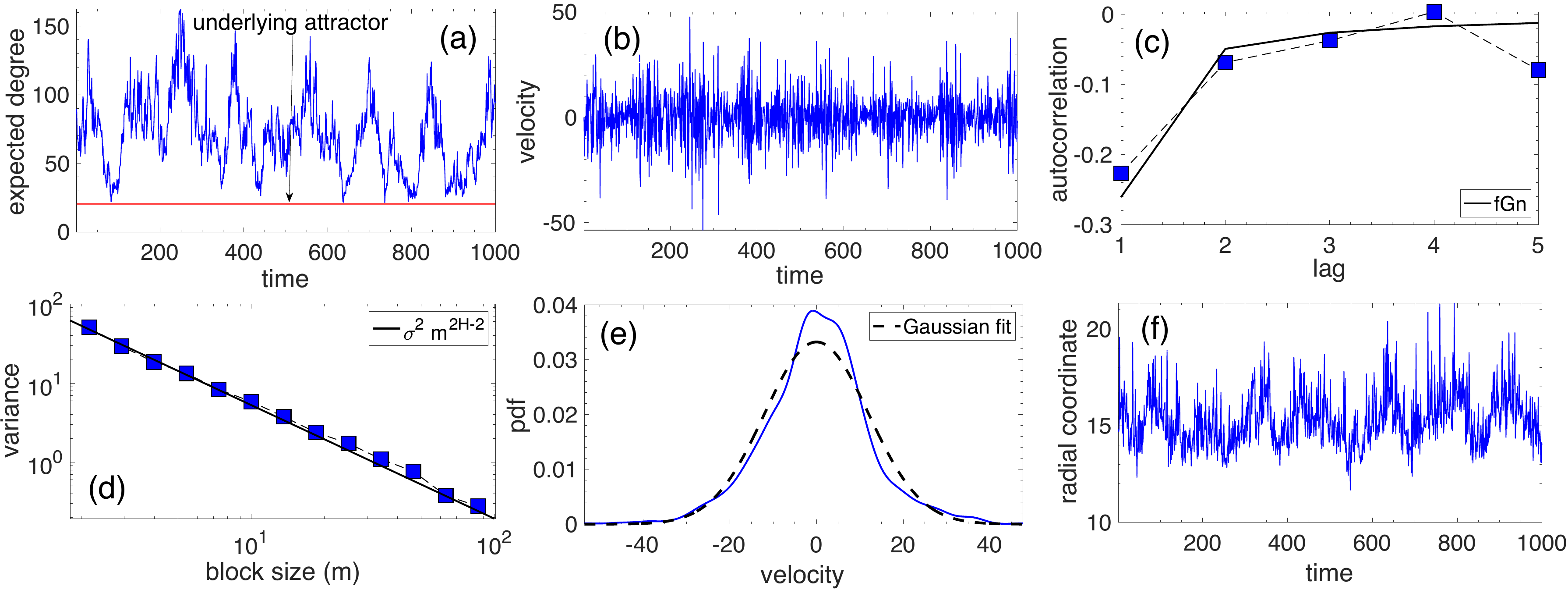}
\caption{Same as in Fig.~\ref{deg_s1}, but for the expected degree of the node in Fig.~\ref{sim_cm} (CM). The node's hidden degree is $\kappa=20.6$ (red line in~(a)). The Hurst exponents for the degree and radial trajectories are respectively $0.28$ and $0.04$.}
\label{deg_cm}
\end{figure*}

\begin{figure*}
\centering
\includegraphics[width=16.5cm]{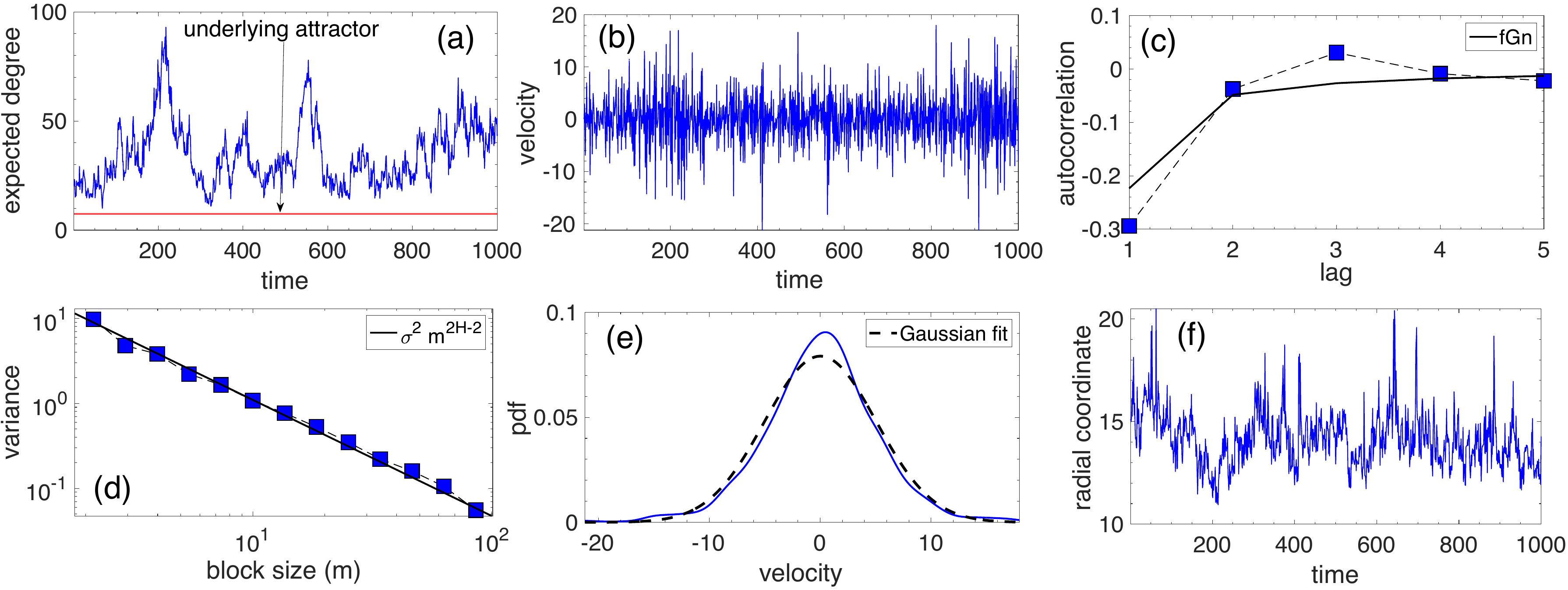}
\caption{Same as in Fig.~\ref{deg_s1}, but for the expected degree of the node in Fig.~\ref{sim_rg} (RGs). The node's hidden degree is $\kappa=7.4$ (red line in~(a)). The Hurst exponents for the degree and radial trajectories are respectively $0.32$ and $0.18$.}
\label{deg_rg}
\end{figure*}

Finally, Figs.~\ref{deg_s1}-\ref{deg_rg} show corresponding results for the expected degree trajectories. In the geometric version of the model, these trajectories also possess similar characteristics as in real networks (Fig.~\ref{deg_s1}). As mentioned in the main text, degree anti-persistence weakens in non-geometric networks (Figs.~\ref{deg_cm}, \ref{deg_rg}). Further, we observe that the degree trajectories can significantly deviate from their underlying attractors, which is not the case in RHGs (cf. Figs.~\ref{deg_s1}-\ref{deg_rg}(a)). Figure~\ref{geom_effect}(a) in the main text shows the distributions of Hurst exponents across all expected degree trajectories in the considered synthetic networks.


%

\end{document}